\documentclass[10pt,twocolumn]{article}
\usepackage[top=50pt,bottom=70pt,left=40pt,right=40pt]{geometry}

\usepackage{cite}

\usepackage[table,xcdraw]{xcolor}
\usepackage{tabularx,booktabs}
\newcolumntype{C}{>{\centering\arraybackslash}X} % centered version of "X" type
\newcolumntype{b}{X}
\newcolumntype{s}{>{\hsize=.5\hsize}X}
\newcolumntype{v}{>{\hsize=.3\hsize}X}
\usepackage{graphicx}
\usepackage[colorinlistoftodos]{todonotes}
\usepackage[colorlinks=true, allcolors=blue]{hyperref}
\usepackage{caption}
\usepackage{latexsym}
\usepackage{enumitem}
\usepackage{amssymb}
\usepackage{subcaption}
\usepackage{mathtools}
\usepackage{multirow}
\usepackage{float}
\usepackage[linesnumbered,ruled,vlined]{algorithm2e} 
\usepackage{cprotect}
\usepackage{xspace}
\usepackage{amsmath} % for the "align" and "align*" environments
\usepackage{soul}
\usepackage{hyphenat}
\usepackage{epstopdf}
\usepackage{threeparttable}
\usepackage{pifont}% http://ctan.org/pkg/pifont
\usepackage{textcomp}
\usepackage[toc,page]{appendix}

\newcommand{\xmark}{\ding{55}}%

\DeclarePairedDelimiter{\ceil}{\lceil}{\rceil}

\DeclareMathOperator*{\argmax}{argmax}

\usepackage{tikz}
\usepackage{tkz-tab}
\usetikzlibrary{automata,arrows,positioning,calc}
\usetikzlibrary{shapes,snakes}

\def\Plus{\texttt{+}}

\DeclarePairedDelimiter\abs{\lvert}{\rvert}

\makeatletter
\newcommand{\removelatexerror}{\let\@latex@error\@gobble}

\newcommand{\newverold}[1]{#1}
\newcommand{\newver}[1]{#1}

\makeatother

\begin{document}
	
	%\title{Effects of Dynamic Channel Bonding on Non-Spatially-Constrained High Density WLANs}	
	\title{\newver{Dynamic Channel Bonding in Spatially Distributed High-Density WLANs}}
	% Insights on Dynamic Channel Bonding in High Density WLANs \textcolor{red}{[Hem de trobar un títol més xulo. Lo d'insights no em convenç. I que surti Hidden \& Exposed Nodes al títol?]} \sergio{[Ok, podem ficar els exposed i hidden]} \textcolor{red}{No, tampoc en convenç... ja trobarem alguna cosa} \sergio{[Proposta 2: Effects of Dynamic Channel Bonding on Non-Spatially-Constrained High Density WLANs]}
	
	\author{Sergio~Barrachina-Mu\~noz, Francesc~Wilhelmi, Boris~Bellalta}
	
	\date{}
	
	\maketitle
		
	\begin{abstract}

		In this paper\newver{,} we discuss the effects on throughput and fairness of dynamic channel bonding (DCB) in spatially distributed \newver{high-density} wireless local area networks (WLANs). First, we present an analytical framework based on \newver{continuous-time} Markov networks (CTMNs) for depicting the \newver{behavior of different} DCB policies in spatially distributed scenarios, where nodes are not required to be within the carrier sense range of each other. Then, we assess the performance of DCB in \newver{high-density} IEEE \newver{802.11ac/ax} WLANs by means of simulations. \newver{We show that there may be critical interrelations among nodes in the spatial domain} -- even if they are located outside the carrier sense range of each other -- in a \textit{chain reaction} manner. Results also \newver{reveal} that, while always selecting the widest available channel normally maximizes the individual long-term throughput, it often generates unfair \newver{situations} where other WLANs starve. Moreover, we show that there are scenarios where DCB with stochastic channel width selection improves the latter approach both in terms of individual throughput and fairness. It follows that there is not a unique \newver{optimal DCB policy} for every case. Instead, smarter bandwidth adaptation is required in the challenging scenarios of next-generation WLANs.
		
		%In this paper we discuss the effects on throughput and fairness of dynamic channel bonding (DCB) in spatially distributed high density (HD) wireless local area networks (WLANs). First, we present an analytical framework based on continuous time Markov networks (CTMNs) for depicting the phenomena given when applying different DCB policies in spatially distributed scenarios, where nodes are not required to be within the carrier sense range of each other. Then, we assess the performance of DCB in HD IEEE 802.11ax WLANs by means of simulations. Regarding spatial distribution, we show that there may be critical interrelations among nodes - even if they are located outside the carrier sense range of each other - in a chain reaction manner. Results also show that, while always selecting the widest available channel normally maximizes the individual long-term throughput, it often generates unfair scenarios where other WLANs starve. Moreover, we show that there are scenarios where DCB with stochastic channel width selection improves the latter approach both in terms of individual throughput and fairness. It follows that there is not a unique DCB policy that is optimal for every case. Instead, smarter bandwidth adaptation is required in the challenging scenarios of next-generation WLANs.
		
	\end{abstract}

	%-------------------------
	%-------------------------
	%--INTRODUCTION --------
	%-------------------------
	%-------------------------
	\section{Introduction}\label{sec:introduction}
	
	% What is and evolution of WLANs (WLANs are nice) and rising problems
	Wireless local area networks (WLANs), with IEEE 802.11 as the most widely used standard, are a cost-efficient solution for wireless Internet access that can satisfy most of the current communication requirements in domestic, public, and business scenarios. However, the scarcity of the frequency spectrum in the industrial, scientific and medical (ISM) radio bands, the increasing throughput demands given by new hungry-bandwidth applications, and the heterogeneity of current wireless network deployments give rise to substantial complexity. Such issues gain importance in dense WLAN deployments, leading to multiple partially overlapping scenarios and coexistence problems. 
	
	% CA and DCB to the rescue
	In this regard, two main approaches \newver{to} optimizing the scarce resources of the frequency spectrum are being deeply studied in the context of WLANs: channel allocation (CA) and channel bonding (CB). % Difference between CA and CB
	While CA refers to the action of allocating the potential transmission channels (i.e., \newver{both the primary and secondary channels}) for a WLAN \newver{or group} of WLANs, CB is \newver{the} technique whereby nodes are allowed to use a contiguous set of idle channels for transmitting in larger bandwidths, \newver{thus} potentially achieving \newver{a} higher throughput.
	
	% We focus on DCB
	This paper focuses on CB, which was \newver{first} introduced in the IEEE 802.11n \newver{(11n)} amendment \newver{by allowing two 20 MHz basic channels to be aggregated into a 40 MHz channel}. Newer amendments like IEEE 802.11ac (11ac) extend the number of basic channels that can be aggregated up to 160 MHz channel widths. It is expected that IEEE 802.11ax (11ax) will boost the use of wider channels \cite{bellalta2016ieee}.
	% SCB vs DCB
	Nonetheless, due to the fact that using wider channels increases the contention and interference among nodes, undesirable lower performances may be experienced when applying static channel bonding (SCB), \newver{especially} in \newver{high-density} WLAN scenarios. To mitigate such a negative effect, dynamic channel bonding (DCB) policies are used to select the bandwidth in a more flexible way based on the instantaneous spectrum occupancy. A well-known example of DCB policy is \textit{always-max} (AM)\footnote{Some papers in the literature \newver{indistinctively} use the terms DCB and AM. In this paper\newver{,} we notate AM as \newver{a} special case of DCB.}  \cite{bellalta2016interactions, park2011ieee}, where transmitters select the widest channel found idle when the backoff counter terminates.
	% Need of the paper
	To the best of our knowledge, the works in the literature assessing the performance of DCB \newver{only study} the SCB and AM policies, while they also assume fully overlapping scenarios where all the WLANs are within the carrier sense range of the others \cite{bellalta2014channel, sv2016performance, bellalta2016interactions, faridi2016analysis}. Therefore, there is an important lack of insights on the performance of CB in more realistic WLAN scenarios, where such a condition usually does not hold. 
	
	% What we do new in this paper I: non fully-overlap + DCB probabilistic
	\newverold{With this work, we aim to extend the state of the art by providing new insights on the performance of DCB under saturation regimes in WLAN scenarios that are not required to be fully overlapping; where the effect of carrier sense and communication ranges play a crucial role due to spatial distribution interdependencies. Namely, \newver{the operation of a node} has a direct impact on the nodes inside its carrier sense range, which in turn may affect nodes located outside such range in complex and hard to prevent ways. Besides, we assess different DCB policies, including \newver{a} stochastic approach that selects the channel width randomly.}
	
	% What we do in this paper II: SFCTMN + toy scenarios
	\newver{In order to evaluate different DCB policies, we first introduce the Spatial-Flexible Continuous Time Markov Network (SFCTMN), an analytical framework based on \newver{continuous-time} Markov networks (CTMNs). This framework is useful for describing the different phenomena that occur in WLAN deployments when considering DCB in spatially distributed scenarios. In this regard, we analytically depict such complex phenomena by means of illustration through several toy scenarios.}
	% What we do in this paper III: simulations in 11axHDWLANSim + results
	\newver{Finally, we evaluate the performance of the proposed policies in large \newver{high-density} 11ax WLAN scenarios by means of simulations using \texttt{Komondor}, \newver{a particular release (v1.0.1) of the Komondor~\cite{barrachina2019komondor} wireless networks simulator.}\footnote{All of the source code of SFCTMN and \texttt{Komondor} is open, encouraging sharing of algorithms between contributors and providing the ability for people to improve on the work of others under the GNU General Public License v3.0. The repositories can be found at \url{https://github.com/sergiobarra/SFCTMN} and \url{https://github.com/wn-upf/Komondor}, respectively.} We find that, while AM is normally the best DCB policy for maximizing the individual long-term throughput of a WLAN, it may generate unfair situations where some other WLANs starve. In fact, there are cases where less aggressive policies like stochastic channel width selection improve AM both in terms of individual throughput and fairness. This leads to the need of boosting wider channels through medium adaptation policies.}
	% Contributions
	The contributions of this paper are as follows:
	\begin{enumerate}
		
		% Insight non-fully-overlaping
		\item Novel insights on the effects of DCB in \newver{high-density scenarios}. We depict the complex interactions given in spatially distributed deployments -- i.e., considering path loss, signal-to-interference-plus-noise ratio (SINR) and clear channel assessment (CCA) thresholds, co-channel and adjacent channel interference, etc. -- and discuss the influence that nodes have \newver{among} them. % In this regard, we show that AM is not always the optimal policy and that different networks may require different DCB policies. This leads to the need of local policy adaptation.
		
		% Definition of policies
		\item Generalization of DCB policies including \textit{only-primary} (i.e., selecting just the primary channel for transmitting), SCB, AM, and \textit{probabilistic uniform} (PU) (i.e., selecting the channel width stochastically).% We show that WLAN scenarios implementing any combination of DCB policies can be characterized by a CTMN with specific transition probabilities. This allows us to analytically compare the behavior of these policies.
		
		% Extension of the algorithm
		\item \newverold{Algorithm for modeling WLAN scenarios with CTMNs that extends the one presented in \cite{faridi2016analysis}. Such an extension allows us capturing non-fully overlapping scenarios, taking into consideration spatial distribution implications. Moreover, this algorithm allows us to model any combination of DCB policies in a network.}
		%Moreover, the presented algorithm allows us to analytically compare the behavior of any combination of DCB policies.}
		
		% Evaluation		
		\item Performance evaluation of the presented DCB policies in \newver{high-density} WLAN scenarios by means of simulations. The selected physical (PHY) and medium access control (MAC) parameters are representative of single user (SU) transmissions in 11ax WLANs. %We show that AM does not always maximize the individual throughput, but more flexible policies like PU are required. Besides, we show that policy learning and/or adaptation is required on a per-WLAN basis in order to maximize both the individual throughput and system fairness.
		
	\end{enumerate}
	
	% Remainder structure
	%\newverold{The remainder of this article is organized as follows. In Section \ref{sec:related_work}, we provide related work on DCB. Then, in Section \ref{sec:system_model} we describe the system model and assumptions considered throughout the article. The SFCTMN framework for analytically modeling WLAN scenarios with CTMNs is detailed in Section \ref{sec:ctmn_model}. We use SFCTMN to depict the interactions in frequency and space in Section \ref{sec:interactions} by means of several toy scenarios. The performance results gathered with \texttt{Komondor} for the different DCB policies in HD WLANs is assessed in Section \ref{sec:evaluation}. We conclude with some final remarks and future work at Section \ref{sec:conclusions}.}
	
	%-------------------------
	%-------------------------
	%-------------------------
	%-------------------------	

	\section{Related work} \label{sec:related_work}
	
	% Introduction to literature on optimizing WNs. Real testbeds 
	Several works in the literature assess the performance of CB by means of analytical models, simulations or testbeds. Authors in \cite{deek2011impact, arslan2010auto} experimentally analyze SCB in IEEE 802.11n WLANs and show that the reduction of Watt/Hertz when transmitting in larger channel widths causes lower SINR at the receivers. This lessens the coverage area consequently and increases the probability of packet losses due to the accentuated vulnerability to interference. Nonetheless, they also show that DCB can provide significant throughput gains when such issues are palliated by properly adjusting the transmission power and data rates.
	
	% 11ac i 11ax accentuen els pros i els cons. SINR pot millorar amb distancies curtes i MIMO
	\newver{The advantages and drawbacks of CB are accentuated with the 11ac and 11ax amendments since larger channel widths are allowed (up to 160 MHz). Nevertheless, it is important to emphasize that in the dense and short-range WLAN scenarios expected in the coming years \cite{bellalta2016ieee}, the issues concerning low SINR values may be palliated. The main reason lies in the shorter distances between transceiver and receiver, and the usage of techniques like spatial diversity multiple-input multiple-output (MIMO)\cite{deek2013joint}.} \newver{An empirical study on CB in 11ac is followed in \cite{zeng2014first}, where authors show that throughput increases by bonding channels.}
	% simulations
	\newver{By means of simulations, authors in \cite{gong2011channel, park2011ieee} assess the performance of DCB in 11ac WLANs, resulting in significant throughput gains. Still, they also corroborate that these gains are severely compromised by the activity of overlapping wireless networks.}

	% Analytical models and studies
	\newver{There are other works in the literature that follow an analytical approach for assessing the performance of CB. For instance, authors in \cite{bellalta2014channel} analytically model and evaluate the performance of CB in short-range 11ac WLANs, proving significant performance boost in presence of low to moderate external interference. In \cite{bellalta2016interactions}, authors show that CB can provide significant performance gains even in high-density scenarios, though it may also cause unfairness.}
	%Likewise, authors in \cite{faridi2016analysis} use a CTMN-based model to explain key properties of DCB such as the sensitivity to the backoff and transmission time distributions, or the high switching times between different dominant states. 
	\newver{Non-saturation regimes are considered in \cite{kim2017throughput, barrachina2019tooverlap}, where authors propose an analytical model for the throughput performance of CB in 11ac/11ax WLANs under both saturated and non-saturated traffic loads. An analytical framework to study the performance of opportunistic CB where 11ac users coexist with legacy users is presented in \cite{han2016performance}.}
	%Similarly, authors in \cite{joshi2012channel} assess the performance of CB in the context of opportunistic spectrum access via an analytical model and conclude that CB is generally beneficial where there is low primary users activity.
	\newver{Recently, an analytical model based on renewal theory showed that 11ac/11ax DCB can improve throughput even in the presence of legacy users \cite{khairy2018renewal}. Literature on CTMN models for DCB WLANs is further discussed in Section \ref{sec:ctmn_model}.}
	
	% Solutions
	\newver{As for particular CB solutions or algorithms, an intelligent scheme for jointly adopting the rate and bandwidth in MIMO 11n WLANs is presented in \cite{deek2013joint}. Testbed experiments show that such scheme (ARAMIS) accurately adapts to a wide variety of channel conditions with negligible overhead and achieving important performance gains}.
	%In \cite{moscibroda2008load}, authors propose an spectrum-distribution algorithm where access points (APs) adjust both the primary channel and width to match their traffic loads. They show that the spectrum utilization can be substantially improved when allocating more bandwidth to APs with high traffic load.
	\newver{A DCB protocol with collision detection is presented in \cite{huang2016dynamic}, in which a node gradually increases the transmission bandwidth whenever new narrow channels are found available.}
	A stochastic spectrum distribution framework accounting for WLANs demand uncertainty is presented in \cite{nabil2017adaptive}, showing better performance compared to the naive allocation approach. Authors in \cite{kai2017channel} show that the maximal throughput performance can be achieved with DCB under the CA scheme with the least overlapped channels among WLANs.
	\newver{A dynamic bandwidth selection protocol for 11ac WLANs is proposed in \cite{chen2018dbs} to prevent the carrier sensing decreasing and outside warning range problems. In \cite{khairy2018renewal} authors propose a heuristic primary channel selection for maximizing the throughput of multi-channel users. Finally, \cite{byeon2018reconn} proposes a prototype implementation for commercial 11ac devices showing up to 1.85x higher throughput when canceling time-domain interference.}

	% Paragraph: Que aporta aquest paper vs estat de l'art
	% Dues coses principalment: non-spatially-constrained i DCB policies.
	\newverold{We believe that this is the first work providing insights into the performance of DCB in spatially distributed scenarios, where the effect of partially overlapping nodes plays a crucial role due to the spatial interdependencies. We also provide an algorithm for generating the CTMNs to model such kind of scenarios. Besides, we assess the performance of different DCB policies, including a novel stochastic approach, and show that always selecting the widest available channel may be sub-optimal in some scenarios.}
	
	%-------------------------
	%-------------------------
	%-------------------------
	%-------------------------
	
	\section{System model under consideration} \label{sec:system_model}
	
	% Intro to the Section
	In this section, we first depict the notation regarding channelization that is used throughout this article. We also define the DCB policies that are studied and provide a general description of the carrier sense multiple access with collision avoidance (CSMA/CA) operation in IEEE 802.11 WLANs. Finally, we expose the main assumptions considered in the presented scenarios.
	%-------------------------
	%-------------------------

	\subsection{Channelization}
	
\begin{figure}[t]
		\centering
		\includegraphics[width=0.4\textwidth]{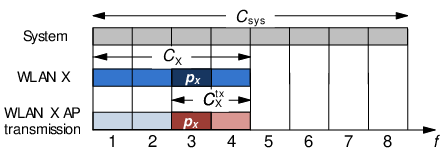}
		\caption{Channel access notation.}    
		\label{fig:channel_allocation_notation}
	\end{figure}

\begin{figure}[t]
		\centering
		\includegraphics[width=0.48\textwidth]{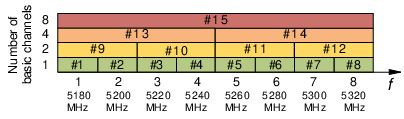}
		\caption{Simplified channelization of 11ac and 11ax.}    
		\label{fig:channel_allocation_ac}
	\end{figure}

\begin{figure*}[ht]
		\centering
		\includegraphics[width=1\textwidth]{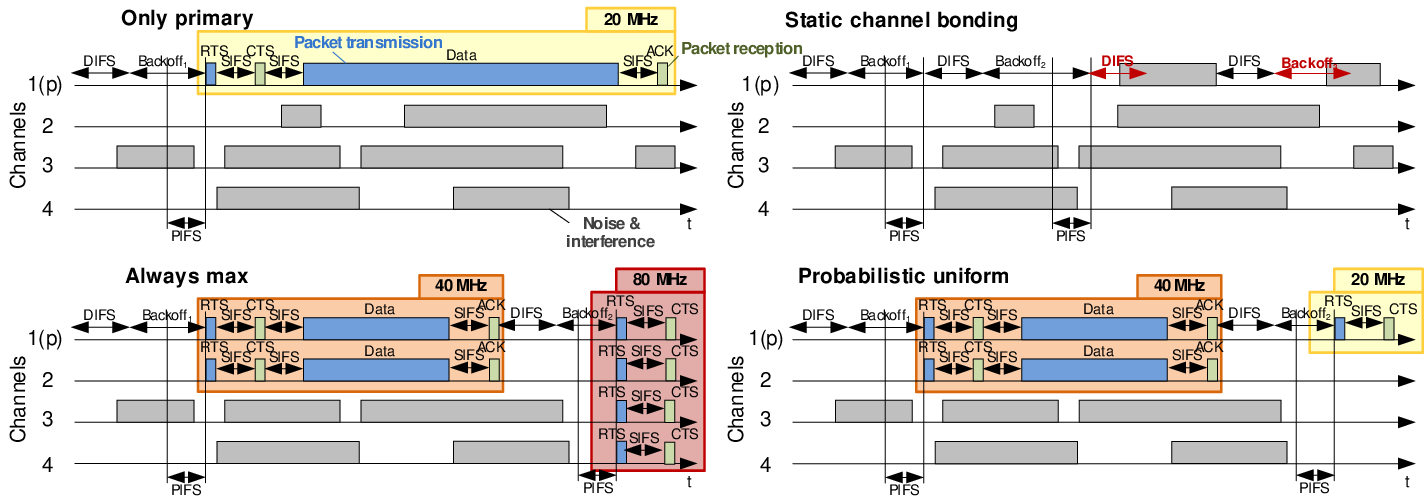}
		\caption{\newverold{CSMA/CA temporal evolution of a node operating under different DCB policies in \newver{an} 11ax channelization scheme. The DIFS and backoff in red represent that the sensed interference in the primary channel forces \newver{resetting} the backoff procedure. While the legacy packets (RTS, CTS, and ACK) duration is the same no matter the bandwidth, the data duration is clearly reduced when transmitted in 40 MHz.}}
		\label{fig:dcb_dcf}
	\end{figure*}

	% Example of channel access notation
	Let us discuss the example shown in Figure \ref{fig:channel_allocation_notation} for introducing the channel access terminology and facilitating further explanation. In this example, the system channel $C_\text{sys}$ counts with $N_\text{sys} = 8$ basic channels and WLAN X is allocated with the channel $C_\text{X} = \{1,2,3,4\}$ and primary channel $p_\text{X} = 3$. Note that in this particular example the AP of WLAN X does not select the entire allocated channel, but a smaller one, i.e.,   $C_\text{X}^{\text{tx}} = \{3,4\} \subseteq C_\text{X}$. Two reasons may be the cause: \textit{i}) basic channels 1 and/or 2 are detected busy at the end of the backoff, or \textit{ii}) the DCB policy determines not to pick them. More formally, the \newver{definitions} of the channelization terms used throughout the paper are as follows:

	\begin{itemize}
		
		\setlength\itemsep{0.2em}
		
		\item \textbf{Basic channel $\boldsymbol{c}$}: the frequency spectrum is split into basic channels of width $|c| = 20$ MHz.
		
		\item \textbf{Primary channel $\boldsymbol{p_w}$}: \newver{a} basic channel with different roles depending on the node state. All the nodes belonging to the same WLAN $w$ must share the same primary channel $p_w$. Essentially, this channel is used to \textit{i}) sense the medium for decrementing the backoff when the primary channel's frequency band is found free, and \textit{ii}) listening to control and data packets. %Hence, the primary channel is always used when transmitting and receiving packets.
		
		\item \textbf{Channel $\boldsymbol{C}$}: a channel $C = \{c_1, c_2, ..., c_N\}$ consists of a contiguous set of $N$ basic channels. The width (or bandwidth) of a channel is $N|c|$.
		
		\setlength\itemsep{0.2em}
		
		\item \newver{\textbf{Channelization scheme $\boldsymbol{\mathcal{C}}$}: the set of channels that can be used for transmitting is determined by the channel access specification and the system channel ($C_{\text{sys}}$), whose bandwidth is given by $N_\text{sys}|c|$. Namely, all the nodes in the system must transmit in some channel included in $\mathcal{C}$. A simplified version of the channelization considered in the 11ac and 11ax standards is shown in Figure \ref{fig:channel_allocation_ac}, $\mathcal{C}=\{\{1\}, \{2\},...,\{1,2\},\{3,4\},...,\{1,2,3,4\},...,\{1,2,...,8\}\}$.}
		
		\item \textbf{WLAN's allocated channel $\boldsymbol{C_w}$}: nodes in a WLAN $w$ must transmit in a channel contained in $C_w \in \mathcal{C}$. Different WLANs may be allocated with different primary channels and different available channel widths.%We use $N_\text{w}$ to denote the number of basic channels allocated for WLAN $w$.
		
		\item \textbf{Transmission channel $\boldsymbol{C_n^\text{tx}}$}: a node $n$ belonging to a WLAN $w$ has to transmit in a channel $C^{\text{tx}}_n \subseteq C_w \in \mathcal{C}$, which will be given by \textit{i}) the set of basic channels in $C_w$ found idle by node $n$ at the end of the backoff ($C_n^\text{free}$),\footnote{Note that, in order to include secondary channels for transmitting, a WLAN must listen them free during at least a point coordination function interframe space (PIFS) period before the backoff counter terminates as shown in Figure \ref{fig:dcb_dcf}. While such PIFS condition is not considered in the SFCTMN framework for the sake of analysis simplicity, the \texttt{Komondor} simulator does.} and \textit{ii}) the implemented DCB policy.
		
	\end{itemize}

	%-------------------------
	%-------------------------
	
	\subsection{CSMA/CA operation in IEEE 802.11 WLANs}
	
	% Backoff generation
	According to the CSMA/CA operation, when a node $n$ belonging to a WLAN $w$ has a packet ready for transmission, it measures the power sensed in the frequency band of $p_w$. Once the primary channel has been detected free, i.e., the power sensed by $n$ at $p_w$ is smaller than its CCA threshold, the node starts the backoff procedure by selecting a random initial value of $\text{BO} \in [0,\text{CW}-1]$ time slots of duration $T_e$. The contention window is defined as $\text{CW} = 2^b \text{CW}_\text{min}$, where $b \in \{0,1,2,...,m\}$ is the backoff stage with maximum value $m$, and $\text{CW}_\text{min}$ is the minimum contention window. When a packet transmission fails, $b$ is increased by one unit, and reset to 0 when the packet is acknowledged.
	
	% Backoff procedure
	After computing $\text{BO}$, the node starts decreasing the backoff counter while sensing the primary channel. Whenever the power sensed by $n$ at $p_w$ is higher than its CCA, the backoff is paused and set to the nearest higher time slot until $p_w$ is detected free again, at which point the countdown is resumed. When the backoff timer reaches zero, the node selects the transmission channel $C_n^{\text{tx}}$ based on the set of idle basic channels $C_n^\text{free}$ and on the DCB policy.
	% RTS/CTS, data packet, ack
	The selected transmission channel is then used throughout the whole packet exchanges involved in a data packet transmission between the transceiver and receiver. Namely, \newver{request to send (RTS) -- used for notifying the selected transmission channel -- clear to send (CTS), and acknowledgment (ACK)} packets are also transmitted in $C_n^{\text{tx}}$. Likewise, any other node that receives an RTS in its primary channel with enough power to be decoded will enter in \newver{network allocation vector (NAV)} state, which is used for deferring channel access and avoiding packet collisions (especially those caused by hidden node situations).
	
	\subsection{DCB policies}
	
	The DCB policy determines \newver{the} transmission channel a node must pick from the set of available ones. When the backoff terminates, any node belonging to a WLAN $w$ operates according to the DCB policy as follows:
	
	\begin{itemize}
		
		\item \textbf{Only-primary (OP) or single-channel}: \newver{pick} just the primary channel for transmitting if it is found idle.
		
		\item \textbf{Static channel bonding (SCB)}: exclusively \newver{pick} the full channel allocated in its WLAN when found free. Namely, nodes operating under SCB cannot transmit in channels different than $C_w$.
		
		\item \textbf{Always-max (AM)}: \newver{pick} the widest possible channel found free in $C_w$ for transmitting.
		
		%% NOTE: posibles critiques per part dels reviewers al fet d'utilitzar probabilistic uniform i no una altra versió 
		\item \textbf{Probabilistic uniform (PU)}: \newver{pick} with same probability any of the possible channels found free inside the allocated channel $C_w$.
		
		% picks the transmitting channel depending on the idle basic channels and on a given probability density function. In this work we study a particular version called probabilistic uniform (PU), which picks with same probability any of the possible available transmission channels found free. 
		
	\end{itemize}	
	
	% Example of CSMA/CA discussion
	For the sake of illustration, let us consider the example shown in Figure \ref{fig:dcb_dcf}, which shows the evolution of a node implementing different DCB policies. In this example, a node is allowed to transmit in the set of basic channels $C_w=\{1(p), 2, 3, 4\}$, where $p_w=1$ is the primary channel. While OP picks just the primary channel, the rest of policies try to bond channels in different ways. 	
	% Actual CB policies
	In this regard, SCB is highly inefficient in scenarios with partial interference. In fact, no packets can be transmitted with SCB in this example since the basic channel $\{3\} \in C_w$ is busy during the PIFS durations previous to the backoff terminations. However, more flexible approaches like AM and PU are able to transmit more than one \newver{frame} in the same period of time. On the one hand, AM adapts in an aggressive way to the channel state. Specifically, it is able to transmit in 40 and 80 MHz channels at the end of the first and second backoff, respectively. On the other hand, the stochastic nature of PU makes it more conservative than AM. In the example, the node could transmit in 1 or 2 basic channels with \newver{the} same probability (1/2) at the end of the first backoff. Likewise, after the second backoff, a channel composed of 1, 2 or 4 basic channels could be selected with probability (1/3).

	\subsection{Main assumptions}
	
	In this paper, we present results gathered via the SFCTMN framework based on CTMNs, and also via simulations through the \texttt{Komondor} wireless network simulator. While in the latter case we are able to introduce more realistic implementations of the 11ax amendment, in the analytical model we use relaxed assumptions for facilitating \newver{subsequent} analysis. This subsection depicts the general assumptions considered in both cases.
	
	\begin{enumerate}
		
		%% NOTE: cal justificar propagation delay. Carrier sense en 5 GHz es petita.
		\item \newverold{\textbf{Channel model}: signal propagation is isotropic. Also, the propagation delay between any pair of nodes is considered negligible because of the small carrier sense in the above 1GHz ISM bands where WLANs operate. Besides, the transmission power is divided uniformly among the basic channels in the selected transmission channel. We also consider an adjacent channel interference model that replicates $\text{P}_\nu$ of the power transmitted per Hertz into the two basic channels that are contiguous to the actual transmission channel.}
		
		\item \textbf{Packet errors}: a packet is lost if \textit{i}) the power of interest received at the receiver is less than its CCA, \textit{ii}) the SINR ($\gamma$) perceived at the receiver does not accomplish the capture effect (CE), i.e., $\gamma <$ CE, or \textit{iii}) the receiver was already receiving a packet. In the latter case, the decoding of the first packet is ruptured only if the CE is no longer accomplished because of the interfering transmission. 
		% Infinite retransmissions
		We assume an infinite maximum number of retransmissions per packet, whose effect is negligible in most of the cases because of the small probability of retransmitting a data packet more than few times \cite{chatzimisios2004performance}.
		
		\item \textbf{Modulation and coding scheme (MCS)}: the MCS index used by each WLAN is the highest possible according to the SINR perceived by the receiver, and it is kept constant throughout all the simulation. \newverold{We assume that the MCS selection is designed to keep the packet error rate constant and equal to $\eta = 0.1$ given that static deployments are considered.}\footnote{\newverold{In 802.11 devices, given a minimum receiver sensitivity and SINR table, a maximum decoding packet error rate of 10\% is usually tried to be guaranteed when selecting the MCS index.}}
		\newverold{Note that the $\eta$ value is only considered if none of the three possible causes of packet error explained in the item above are given.}
		
		% NOTE: no he trobat cap cita per dir q DL es el tipus de trafic mes comu a les WLAN
		\item \textbf{Traffic}: downlink traffic is considered. 
		%Namely, while RTS and data packets are transmitted by the APs, CTS and ACK packets are transmitted by the STAs. 
		In addition, we assume a full-buffer mode where APs always have backlogged data pending for transmission. 
		
	\end{enumerate}

	%-------------------------
	%-------------------------
	%-------------------------
	%-------------------------
	
	% NOTE: pending for revision. 
	\section{The CTMN model for WLANs}	\label{sec:ctmn_model}
	
	% Several works in the literature use CTMNs for WLANs
	The analysis of CSMA/CA networks through CTMN models was firstly introduced in \cite{boorstyn1987throughput}. Such models were later applied to IEEE 802.11 networks in \cite{bianchi2000performance, faridi2016analysis, bellalta2016interactions, kim2017throughput,bellalta2014channel, kai2017channel, bellalta2017throughput, barrachina2019tooverlap}, among others. Experimental results in \cite{nardelli2012closed, liew2010back} demonstrate that CTMN models, while idealized, provide remarkably accurate throughput estimates for actual IEEE 802.11 systems. A comprehensible example-based tutorial of CTMN models applied to different wireless networking scenarios can be found in \cite{bellalta2014throughput}.
	% But overlapping is required
	\newver{Nevertheless, to the best of our knowledge, works that model DCB through CTMNs study just the SCB and AM policies, while assuming fully overlapping scenarios. Therefore, there is an important lack of insights on more general deployments, where such conditions usually do not hold and interdependencies among nodes may have a critical impact on their performance.} \newver{For instance, an optimal channel allocation algorithm to achieve maximal throughput with DCB was recently presented in~\cite{kai2017bond}. However, this work does not consider the implications of either spatial distribution nor CE.}
	
	% We extend the CTMN to allow spatially flexible scenarios
	In this section, we depict our extended version of the algorithm introduced in \cite{faridi2016analysis} for generating the CTMNs corresponding to spatially distributed WLAN scenarios, which is implemented in the SFCTMN framework. With this extension, as the condition of having fully overlapping networks is no longer required for constructing the corresponding CTMNs, more factual observations can be made.
	
	% WLAN-centric model
	% \textcolor{blue}{In this paper, we follow the WLAN-centric modelling approach presented in \cite{bellalta2016interactions}. This is, we model the aggregate the behavior of all nodes in a WLAN as if they were a single entity, including how they interact with nodes from other WLANs. However, it is worth to mention that since we consider the AP is the only node transmitting per WLAN, there are no differences between node-centric and WLAN-centric approaches, while it allows us to keep the explanations and discussion about how WLANs interact at a network level.} \boris{Jo no diria massa més, a no ser que el reviewer insisteixi}.
	
	%In this paper, since we consider that there is only one transmitter-receiver pair per WLAN (i.e., the AP and one STA), we will simply refer to the WLAN activity as a unit.
	
	\subsection{Implications}
	
	% SFCTMN uses CTMNs. NO BO collisions.
	Modeling WLAN scenarios with CTMNs requires the backoff and transmission times to be exponentially distributed. \newver{It follows that,} because of the negligible propagation delay, the probability of packet collisions between two or more nodes within the carrier sense range of the others is zero. The reason is that two WLANs will never end their backoff at the same time, and therefore they will never start a transmission at the same time either. Besides,
	% CTMNs and DCB
	in overlapping single-channel CSMA/CA networks, it is shown that the state probabilities are insensitive to the backoff and transmission time distributions \cite{liew2010back,salameh2014spectrum}. However, even though authors in \cite{faridi2016analysis} prove that the insensitivity property does not hold for DCB networks, the sensitivity to the backoff and transmission time distributions is very small. Therefore, the analytical results obtained using the exponential assumption offer a good approximation for deterministic distributions of the backoff, data rate, and packet length.
	
	\subsection{Constructing the CTMN}
	
\begin{figure}[t]
		\centering
		\includegraphics[width=0.3\textwidth]{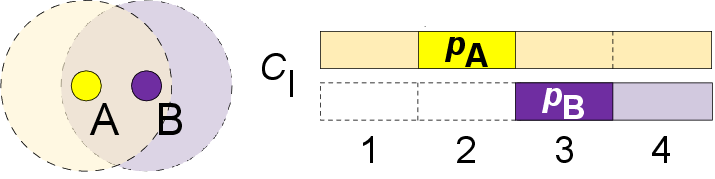}
		\caption{\textit{Scenario I}. WLANs A and B are inside the carrier sense range of each other with potentially overlapping basic channels 3 and 4.}
		\label{fig:scenario_I}
	\end{figure}

	% Toy scenario to introduce the algorithm
	In order to depict how CTMNs are generated, let us consider the toy scenario (\textit{Scenario I}) shown in Figure \ref{fig:scenario_I}, which is composed of two fully overlapping WLANs implementing AM.\footnote{\textit{Scenario I} is selected for conveniently depicting the algorithm. CTMNs corresponding to non-fully overlapping scenarios (e.g., \textit{Scenario III} in Section \ref{sec:interactions}) can be also generated with the very same algorithm.} The channel allocation \newver{employed} in such a scenario can be defined as $C_\text{A} = \{1, 2, 3, 4\}$ with $p_\text{A} = 2$, and $C_\text{B} = \{3, 4\}$ with $p_\text{B} = 3$. That is, there are four basic channels in the system, and the set of valid transmission channels according to the 11ax channel access scheme is $\mathcal{C}_\text{I} = \{\{1\}, \{2\}, \{3\}, \{4\}, \{1,2\}, \{3,4\}, \{1,2,3,4\}\}$ (see Figure \ref{fig:channel_allocation_ac}).
	% Overlapping effects. One transmission at the same time.
	Due to the fact that both WLANs are inside the carrier sense range of each other, their APs could \newver{transmit} simultaneously at any time $t$ only if their transmission channels do not overlap, i.e., $C^{\text{tx}}_\text{A}(t) \cap C^{\text{tx}}_\text{B}(t) = \emptyset$. Notice that slotted backoff collisions cannot occur because their counters decrease continuously in time, and therefore two transmissions \newver{can} be neither started nor finished at the very same time.
	
	\subsubsection{States}
	
	% What is a global state
	A state in the CTMN is defined by the set of WLANs active and the basic channels on which they are transmitting. Essentially, we say that a WLAN is active if it is transmitting in some channel, and inactive otherwise. We define two types of state spaces: the global state space ($\Psi$) and the feasible state space ($\mathcal{S}$).
	
	% Scenario I figure
	\begin{itemize}
		\item \textbf{Global state space}: a global state $\psi \in \Psi$ is a state that accomplishes two conditions: \textit{i}) the channels in which the active WLANs are transmitting comply with the channelization scheme $\mathcal{C}$, and \textit{ii}) all active WLANs transmit inside their allocated channels. That is, $\Psi$ only depends on the particular channelization scheme $\mathcal{C}$ in use and on the channel allocation of the WLANs in the system. In this paper, we assume that every transmission should be made in channels inside $C_\text{sys}$ that are composed of $a = 2^k$ contiguous basic channels, for some integer $k \leq \log_2 N_\text{sys}$, and that their rightmost basic channels fall on multiples of $a$, as stated in the 11ac and 11ax amendments.
		\item \textbf{Feasible state space}: a feasible state $s \in \mathcal{S} \subseteq \Psi$ exists only if each of the active WLANs in such state started their transmissions by accomplishing the CCA requirement derived from the assigned DCB policy. Namely, given a global state space, $\mathcal{S}$ depends only on the spatial distribution and on the DCB policies assigned to each WLAN.
	\end{itemize} 
	
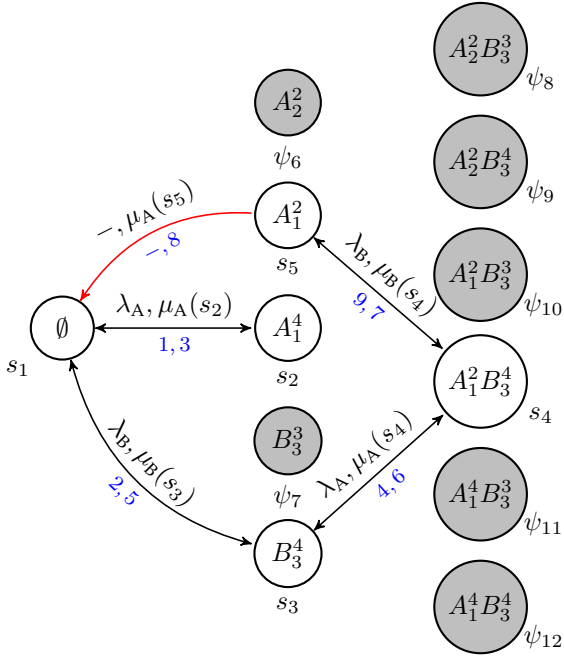
\begin{figure} [t]
		\centering
		\begin{tikzpicture}[<->,>=stealth',shorten >=1pt,auto,node distance=1.5cm,
		semithick]
		\tikzstyle{every state}=[fill=white,draw=black,thick,text=black,scale=1]
		
		\node[state, label=below left:$s_1$]    (S1b)                    {$\emptyset$};
		
		\node[state, label=below:$s_2$]    (S2b)[right of = S1b, xshift=1.5cm]		{$A_{1}^{4}$};
		
		\node[state, label=below:$s_5$]    (S5b)[above of = S2b]		{$A_{1}^{2}$};
		
		\node[state, label=below:$\psi_6$, fill=lightgray]    (S6b)[above of = S5b]		{$A_{2}^{2}$};
		
		\node[state, label=below:$\psi_7$, fill=lightgray]    (S7b)[below of = S2b]		{$B_{3}^{3}$};
		
		\node[state, label=below:$s_3$]    (S3b)[below of = S7b]		{$B_{3}^{4}$};
		
		\node[state, label={[xshift=0.8cm, yshift=-1.3cm]:$s_4$}]    (S4b)[below right of = S2b, xshift=1.5cm, yshift=0.35cm]		{$A_1^2 B_{3}^{4}$};
		
		\node[state, label={[xshift=0.84cm, yshift=-1.3cm]:$\psi_{11}$}, fill=lightgray]    (S11b)[below of = S4b]		{$A_1^4 B_{3}^{3}$};
		
		\node[state, label={[xshift=0.84cm, yshift=-1.3cm]:$\psi_{12}$}, fill=lightgray]    (S12b)[below of = S11b]		{$A_1^4 B_{3}^{4}$};
		
		\node[state, label={[xshift=0.84cm, yshift=-1.3cm]:$\psi_{10}$}, fill=lightgray]    (S10b)[above right of = S2b, xshift=1.5cm, yshift=-0.35cm]		{$A_1^2 B_{3}^{3}$};
		
		\node[state, label={[xshift=0.8cm, yshift=-1.3cm]:$\psi_9$}, fill=lightgray]    (S9b)[above of = S10b]		{$A_2^2 B_{3}^{4}$};
		
		\node[state, label={[xshift=0.8cm, yshift=-1.3cm]:$\psi_8$}, fill=lightgray]    (S8b)[above of = S9b]		{$A_2^2 B_{3}^{3}$};
		
		\path
		
		(S1b) edge     node[sloped, anchor=center, above]{$\lambda_\text{A},\mu_\text{A}(s_2)$} node[sloped, anchor=center, below]{\footnotesize \textcolor{blue}{$1,3$}} (S2b)
		
		(S1b) edge[bend right]    node[sloped, anchor=center, above]{\textcolor{black}{$\lambda_\text{B},\mu_\text{B}(s_3)$}} node[sloped, anchor=center, below]{\footnotesize \textcolor{blue}{$2,5$}} (S3b)
		
		(S1b) edge[bend left, <-, red]    node[sloped, anchor=center, above]{\textcolor{black}{$-,\mu_\text{A}(s_5)$}} node[sloped, anchor=center, below]{\footnotesize \textcolor{blue}{$-,8$}} (S5b)
		
		(S3b) edge     node[sloped, anchor=center, above]{$\lambda_\text{A}, \mu_\text{A}(s_4)$} node[sloped, anchor=center, below]{\footnotesize \textcolor{blue}{$4,6$}} (S4b)
		
		(S5b) edge   node[sloped, anchor=center, above]{$\lambda_\text{B},\mu_\text{B}(s_4)\text{  }$} node[sloped, anchor=center, below]{\footnotesize \textcolor{blue}{$9,7$}} (S4b);

		\end{tikzpicture}
		\caption[]{CTMN of \textit{Scenario I} when applying AM. Circles represent states. All the states are global. Specifically, feasible states are displayed in white, while non-feasible states are gray colored. Two-way transitions are noted with forward and backward rates $\lambda, \mu$, respectively, to avoid cluttering in the figure. The only backward transition is colored in red. The blue pair of numbers beside the transition edges represent the algorithm's discovery order of the forward and backward transitions, respectively.}
		\label{tikz:scenarioI_ctmn}
		
	\end{figure}

	% State notation
	The CTMN corresponding to the toy \textit{Scenario I} is shown in Figure \ref{tikz:scenarioI_ctmn}. Regarding the notation, we represent the states by the most left and most right basic channels used in the transmission channels of each of their active WLANs. For instance, state $s_4 = \text{A}_1^2\text{B}_3^4$ refers to the state where A and B are transmitting in channels $C^{\text{tx}}_\text{A} = \{1,2\}$ and $C^{\text{tx}}_\text{B}=\{3,4\}$, respectively.
	% Example of unfeasible states
	Concerning the state spaces, states $\psi_{6}$, $\psi_{7}$, $\psi_{8}$, $\psi_{9}$, $\psi_{10}$, $\psi_{11}$, $\psi_{12} \notin \mathcal{S}$ are not reachable (i.e., they are global but not feasible) for two different reasons. First, states $\psi_{11}$ and $\psi_{12}$ are not feasible because of the overlapping channels involved. Secondly, the rest of unfeasible states are so due to the fact that AM is applied, thus at any time $t$ that WLAN A(B) finishes its backoff and B(A) is not active, A(B) picks the widest available channel, i.e., $C^{\text{tx}}_\text{A}(t) = \{1,2,3,4\}$ or $C^{\text{tx}}_\text{B}(t) = \{3,4\}$, respectively. Likewise, any time A(B) finishes its backoff and B(A) is active, A(B) picks again the widest available channel, which in this case would be $C^{\text{tx}}_\text{A}(t) = \{1,2\}$ for A and $C^{\text{tx}}_\text{B}(t) = \{3,4\}$ for B if A is not transmitting in its full allocated channel, respectively.
	% Example of backward transition
	Some states such as $s_5 = \text{A}_1^2$ are reachable only via backward transitions. In this case, when A finishes its backoff and B is transmitting in $C^{\text{tx}}_\text{B}(t) = \{3,4\}$ (i.e., $s_3$), A picks just $C^{\text{tx}}_\text{A}(t) = \{1,2\}$ because the power sensed in channels 3 and 4 exceeds the CCA as a consequence of B's transmission. That is, $s_5$ is only reachable through a backward transition from $s_4$, given when B finishes its transmission in state $s_4$.
	
	\begin{algorithm}[htbp]
		
		\cprotect\caption{CTMN generation of spatially distributed DCB WLAN scenarios.}	\label{alg:feasible_states_ctmc_construction}
		$i \coloneqq 1$; \footnotesize \textcolor{white}{.........}\textcolor{darkgray}{\# Index of the last state found}\\
		\normalsize  					% Index of the last state found by the algorithm
		$k \coloneqq 1$; \footnotesize \textcolor{white}{.......} \textcolor{darkgray}{\# Index of the state currently being explored}\\
		\normalsize  							% Index of the state currently being used by the algorithm for discovery
		$s_k \coloneqq \emptyset$; \footnotesize \textcolor{white}{......}\textcolor{darkgray}{\# State currently being explored}\\
		\normalsize  					% State currently being used by the algorithm for discovery
		$\mathcal{S} \coloneqq \{s_k\}$; \footnotesize  \textcolor{darkgray}{\# Set of feasible states}\\ 	% Set of feasible states. Empty state always belong to S
		$Q \coloneqq [\text{ }]$; \footnotesize \textcolor{white}{.......} \textcolor{darkgray}{\# Transition rate matrix}\\					
		
		% Possible forward states
		\footnotesize
		\textcolor{darkgray}{\# Generate the global state space $\Psi$}\\
		\normalsize
		$\Psi \coloneqq $ \footnotesize \texttt{generate\_psi\_space()} \normalsize\;
		
		% While there are known states in S not explored yet
		\While{$s_k \in \mathcal{S}$}{
			
			\ForEach{WLAN X}{
				
				%% If WLAN X is active (present) in current state s_k --->
				%% Set backward transitions to known and unknown states 
				
				\footnotesize \textcolor{darkgray}{\# If WLAN is active in $s_k$}\\
				\normalsize
				
				\uIf{$ \exists a,b \text{ \normalfont{s.t.} } \text{X}_a^b \in s_k$}{
					
					%% For each global state
					\ForEach{$\psi \in \Psi$}{
						
						\footnotesize \textcolor{darkgray}{\# If there exists a \textit{backward} transition}\\
						\normalsize
						\If{$s_k - \text{X}_a^b == \psi$}{
							
							% If destination state NOT known in S, add it
							\If{$\psi \not \in \mathcal{S}$}{
								
								$i \coloneqq i \Plus 1$\;	
								$k' \coloneqq i$\;	
								$S \coloneqq S \cup \psi$\;
							}
							
							\uElse{
								\footnotesize \textcolor{darkgray}{\# Get index of state $\psi$ in $\mathcal{S}$}\\
								\normalsize
								$k' \coloneqq \footnotesize \texttt{get\_index} \normalsize(\varphi^*)$\;
							}
							
							% Add new transition
							\footnotesize
							\textcolor{darkgray}{\# New \textit{backward} transition $s_k \rightarrow s_{k'}$}\\
							\normalsize
							$Q_{k,k'} \coloneqq \mu_\text{X}(s_k)$
						}
					}
				}
				
				%% If wlan is NOT ACTIVE in s_k ---> find forward transitions to other states if any
				%% \Psi_{s_k+X}: set of all the states that differ from s_k only by the participation of
				%% WLAN X.
				
				\footnotesize \textcolor{darkgray}{\# If WLAN is NOT active in $s_k$}\\
				\normalsize
				\uElse{
					
					$\varPhi \coloneqq \emptyset$; \footnotesize \textcolor{white}{.} \textcolor{darkgray}{\# Set of global states reachable from $s_k$}\\
					\normalsize
					$\varPhi^* \coloneqq \emptyset$; \footnotesize \textcolor{white}{.} \textcolor{darkgray}{\# Set of feasible states reachable from $s_k$}\\
					
					% Possible forward states
					\footnotesize
					\textcolor{darkgray}{\# Find possible forward states}\\
					\normalsize
					
					\ForEach{$\psi \in \Psi$}{
						
						\If{$\exists a,b \text{ \normalfont{s.t.} } s_k \Plus \text{X}_a^b == \psi$}{
							
							$\varPhi \coloneqq \varPhi \cup \psi$\;
							
						}
						
					}
					
					\footnotesize
					\textcolor{darkgray}{\# Function $f$ finds feasible states and corresponding transition probabilities according to the DCB policy $\mathcal{D}$ and the basic channels found free}
					\normalsize
					$\mathcal{C}^{\text{free}}_\text{X}(s_k) \coloneqq \{P^{\text{rx}}_\text{X}(s_k,C_\text{X}) < \text{CCA}_\text{X} : 0,1\}$\;
					$\{\varPhi^*, \vec{\alpha}\} \coloneqq f(\mathcal{D},\mathcal{C}^{\text{free}}_\text{X}, \varPhi)$\; \label{alg_line:dcb_policy_range}
					
					%% If WLAN X is able to transmit in n > 0 channels because power sensed in transmission channel is low
					\ForEach{$\varphi^* \in \varPhi^*$}{
						
						%% If state s' not included in S
						\uIf{$\varphi \not \in S$}{
							$i \coloneqq i \Plus 1$\;
							$k'\coloneqq i$\;
							$S \coloneqq S \cup \varphi^*$\;
						}
						
						\uElse{
							\footnotesize \textcolor{darkgray}{\# Get index of state $\varphi^*$ in $\mathcal{S}$}\\
							\normalsize
							$k' \coloneqq \footnotesize \texttt{get\_index} \normalsize(\varphi^*)$\;
						}
						
						\footnotesize
						\textcolor{darkgray}{\# New \textit{forward} transition $s_k \rightarrow s_{k'}$}\\
						\normalsize
						
						$Q_{k,k'} \coloneqq \vec{\alpha}(\varphi^*)\lambda_\text{X}$\;	\label{alg_line:dcb_policy_transition}
					}
				}
			}
			$k \coloneqq k \Plus 1$\;
		}

	\end{algorithm}

	\subsubsection{CTMN algorithm: finding states and transitions}
	
	%% NOTE: I think we should define the algorithm as WLAN-centric or something like this. This would allow to talk about just WLANs, and forget about nodes. This is actually how the algorithm works, but it is not explained yet.
	%% NOTE: problem with the reference to the algorithm.
	% Firstly we have to identify the global state space. Easy
	The first step for constructing the CTMN is to identify the global state space $\Psi$, which is simply composed by all the possible combinations given by the system channelization scheme and the channel allocations of the WLANs. The feasible states in $\mathcal{S}$ are later identified by exploring the states in $\Psi$. Algorithm \ref{alg:feasible_states_ctmc_construction} shows the pseudocode for identifying both $\mathcal{S}$ and the transitions among such states, which are represented by the transition rate matrix $Q$.\footnote{Notice that we use $X_a^b \in s$ to say that a WLAN X transmits in a range of contiguous basic channels $[a,b]$ when the CTMN is in state $s$. With slight abuse of notation, $s - X_a^b$ represents the state where all WLANs that were active in $s$ remain active except for X, which becomes inactive after finishing its packet transmission. Similarly, $s \Plus X_a^b$, represents the state \newver{where} all the active WLANs in $s$ remain active and X is transmitting in the range of basic channels $[a,b]$.}

	% What the algorithm does. Pseudo code explanation.
	% Backward transitions
	Essentially, while there are discovered states in $\mathcal{S}$ that have not been explored yet, for any state $s_k \in \mathcal{S}$ not explored, and for each WLAN X in the system, we determine if X is active or not. If X is active, we then set possible \textit{backward} transitions to already known and not known states. To do so, it is required to fully explore $\Psi$ looking for states where: \textit{i)} other active WLANs in the state remain transmitting in the same transmission channel, and \textit{ii)} WLAN X is not active.

	% Forward transitions
	On the other hand, if WLAN X is inactive in state $s_k$, we try to find \textit{forward} transitions to other states. To that aim, the algorithm fully explores $\Psi$ looking for states where \textit{i}) other active WLANs in the state remain transmitting in the same transmission channel, and \textit{ii}) X is active in the new state as a result of applying the implemented DCB policy ($\mathcal{D}$) as shown in line \ref{alg_line:dcb_policy_range}. It is important to remark that in order to apply such policy, the set of idle basic channels in state $s_k$, i.e., $\mathcal{C}^{\text{free}}_\text{X}(s_k)$, must be identified according to the power sensed in each of the basic channels allocated to X, i.e., $P^{\text{rx}}_\text{X}(s_k,C_\text{X})$, and on its CCA level. \newver{Thereafter}, the transmission channel is selected through the $f$ function, which applies $\mathcal{D}$.	

	% Transition rates
	Each transition between two states $s_i$ and $s_j$ has a corresponding transition rate $Q_{i,j}$. For \textit{forward} transitions, the packet transmission attempt rate (or simply backoff rate) has an average duration $\lambda = 1/(\text{E}[B] \cdot T_\text{slot})$, where $\text{E}[B]$ is the expected backoff duration in time slots, determined by the minimum contention window, i.e., $\text{E}[B] = \frac{\text{CW}_\text{min}-1}{2}$. Furthermore, for \textit{backward} transitions, the departure rate ($\mu$) depends on the duration of a successful transmission, i.e., $\mu = 1/T_\text{suc} = (T_\text{RTS} \Plus T_\text{SIFS} \Plus T_\text{CTS} \Plus T_\text{SIFS} \Plus T_\text{DATA} \Plus T_\text{SIFS} \Plus T_\text{BACK} \Plus T_\text{DIFS} \Plus T_\text{e})^{-1}$, which in turn depends on both the data rate ($r$) given by the selected MCS and transmission channel width, and on the packet length $\text{E}[L_\text{data}]$. In the algorithm, we simply \newver{consider} that the data rate of a WLAN X depends on the state of the system, which collects such information, i.e., $\mu_\text{X}(s)$.
	
	% Forward transition rates may depend on the policy	
	Depending on the DCB policy, different feasible \textit{forward} transitions may exist from the very same state, which are represented by the set $\varPhi^*$. As shown in line \ref{alg_line:dcb_policy_transition}, every feasible \textit{forward} transition rate is weighted by a transition probability vector ($\vec{\alpha}$) whose elements determine the probability of transiting to each of the possible global states in $\varPhi$. Namely, with \newver{a} slight abuse of mathematical notation, the probability to transit to any given feasible state $\varphi^* \in \varPhi$ is  $\vec{\alpha}(\varphi^*)$. As a consequence, $\vec{\alpha}$ must follow the normalization condition $\sum \vec{\alpha}(\varphi^*)=1$.
	
	For the sake of illustration, in the CTMNs of Figures \ref{tikz:scenarioI_ctmn}, \ref{tikz:scenarioII_ctmc} and \ref{tikz:scenario3_ctmn}, states are numbered according to the order in which they are discovered. Transitions between states are also shown below the edges. Note that with SFCTMN, since non-fully overlapping networks are allowed, transitions to states where one or more WLANs may suffer from packet \newver{losses} due to interference are also reachable (see Section \ref{sec:interactions}).
	
	\subsection{Performance metrics}
	
	%% NOTE: discuss local balance may not hold
	% Reason why we can find a solution
	Since there are a limited number of possible channels to transmit in, the constructed CTMN will always be finite. Furthermore, it will be irreducible due to the fact that backward transitions between neighboring states are always feasible. Therefore, a steady-state solution to the CTMN always exists. However, due to the possible existence of one-way transitions between states, the CTMN is not always time-reversible and the local balance may not hold \cite{kelly2011reversibility}. Accordingly, it prevents to find simple product-from solutions to compute the equilibrium distribution of the CTMNs.
	% Explain \pi * Q = 0
	The equilibrium distribution vector $\vec{\pi}$ represents the fraction of time the system spends in each feasible state. Hence, we define $\pi_s$ as the probability of finding the system at state $s$. In order to obtain $\vec{\pi}$ we can use the transition rate matrix $Q$ given the system of equations $\vec{\pi} Q = 0$.
	
	% Use toy scenario as an example
	As an example, for \textit{Scenario I}, considering that its elements are sorted by the discovery order of the states, $\vec{\pi} = (\pi_\emptyset, \pi_{\text{A}_1^4}, \pi_{\text{B}_3^4}, \pi_{\text{A}_1^{2} \text{B}_3^4}, \pi_{\text{A}_1^2})$. Besides, the corresponding transition rate matrix is
	\small\[Q =\left(\begin{array}{cccccc}
	* & \lambda_\text{A} & \lambda_\text{B} & 0 & 0 \\
	\mu_\text{A}(s_2) & * & 0 & 0 & 0 \\
	\mu_\text{B}(s_3) & 0 & * & \lambda_\text{A} & 0 \\
	0 & 0 & \mu_\text{A}(s_4) & * & \mu_\text{B}(s_4) \\
	\mu_\text{A}(s_5) & 0 & 0 & \lambda_\text{B} & * \end{array} \right) \text{,} \]  
	\normalsize where $\lambda_\text{A}$, $\lambda_\text{B}$ and $\mu_\text{A}(s)$, $\mu_\text{B}(s)$ are the packet generation and departure rates in state $s$ of WLANs A and B, respectively. The diagonal elements represented by `*' in the matrix should be replaced by the negative sum of the rest of items of their row, e.g., $Q_{4,4} = -\big(\mu_\text{A}(s_4) \Plus \mu_\text{B}(s_4)\big)$, but for the sake of \newver{illustration} we do not include them in the matrix.
	%% NOTE: beware of the difference between WLAN and node.
	Once $\vec{\pi}$ is computed, estimating the average throughput experienced by each WLAN is straightforward. Specifically, the average throughput of a WLAN $w$ is
	% https://es.wikipedia.org/wiki/Funci%C3%B3n_escal%C3%B3n_de_Heaviside
	\begin{multline*} \label{eq:throughput_ctmn}
	\Gamma_{w} := \text{E}[L]  \bigg(\sum_{s \in \mathcal{S}}^{}\{\gamma_{w}(s) > \text{CE} : 0, 1\}	\mu_w(s)\pi_s \big(1-\eta \big) \bigg) \text{,}
	\end{multline*}
	where $\text{E}[L]$ is the expected data packet length, $\gamma_{w}(s)$ is the SINR perceived by the STA in WLAN $w$ in state $s$, CE is the capture effect, and $\eta$ is the constant packet error probability. The system aggregate throughput is \newver{therefore} the sum of the throughputs of all the WLANs, i.e., $\Gamma := \sum_{w=1}^{M} \Gamma_{w}$. \newver{Besides, in order to evaluate the fairness of a given scenario, we can use $\mathcal{P} := \sum_{w=1}^{M} \log_{10} \Gamma_{w}$ and/or $\mathcal{F} := \big(\sum_{w}^{M} \Gamma_{w}\big)^2/\big(M\sum_{w}^{M} \Gamma_{w}^2\big)$, for proportional fairness and the Jain's Fairness Index, respectively.}
	
	\subsection{Captured phenomena}
	% Main limitations of SFCTMN
	Even though most of the well-known wireless phenomena are captured by SFCTMN, there are some important features that cannot be implemented due to its mathematical modeling nature. Essentially, the main limitations are the inability to capture backoff collisions and the constraints in terms of execution time for medium size networks. Besides, only the overhead of the RTS/CTS packets are considered in SFCTMN, i.e., the time of a successful transmission takes also the transmission time of such packets into account. However, the main purpose of the RTS/CTS mechanism of avoiding hidden nodes is not captured by the generated CTMNs since no packets are actually transmitted. That is, only average performance is captured through states modeled without differentiating from the type of packet being transmitted.
	
	% Why do we need Komondor
	To cope with the \newver{abovementioned} limitations, we make use of a simulator in Section \ref{sec:evaluation}. Basically, when we lose the benefits of analytically modeling the networks, with \texttt{Komondor} we are allowed to simulate large networks and get more realistic insights. A comparison of the features implemented in each tool is shown in Table \ref{table:sfctmn_vs_komondor}.

	%-------------------------
	%-------------------------
	%-------------------------
	%-------------------------
	
	\section{Interactions in frequency and space}	\label{sec:interactions}
	
	% Intro to the section
	In this section, we draw some relevant conclusions about applying different DCB policies in CSMA/CA WLANs by analyzing four representative toy scenarios with different channel allocations and spatial distributions. To that aim, we use the SFCTMN analytical framework \newver{and validate} the gathered results by means of the \texttt{Komondor} wireless network simulator.\footnote{For the sake of saving space, the evaluation setups and corresponding results of the scenarios considered through the paper are detailed in \url{https://github.com/sergiobarra/data_repos/tree/master/barrachina2018performance}}
	
	% Conclusions to be derived from the following scenarios
	In summary, the main outcomes derived from the analytical analyses performed below in this section are: \textit{i}) the feasible system states depend on the DCB policies followed by each of the WLANs, \textit{ii}) maximizing the instantaneous throughput may not be the optimal strategy to maximize the long-term throughput, \textit{iii}) in non-fully overlapping scenarios, cumulative interference and flow starvation may appear and cause poor performance to some WLANs, \textit{iv}) there is not a unique optimal DCB policy. \newver{Note that, otherwise stated, in this paper the optimal DCB policy $\mathcal{D}^*_w$ for WLAN $w$ is the one that maximizes its throughput, i.e., $\mathcal{D}^*_w = \argmax_\mathcal{D} \Gamma_w$.}
	
%	\begin{enumerate}
%		\item \newverold{The feasible system states depend on the DCB policies followed by each of the WLANs.}
%		\item\newverold{Maximizing the instantaneous throughput may not be the optimal strategy to maximize the long-term throughput. \newver{Note that, otherwise stated, in this paper the optimal DCB policy $\mathcal{D}^*_w$ for WLAN $w$ is the one that maximizes its throughput, i.e., $\mathcal{D}^*_w = \argmax_\mathcal{D} \Gamma_w$.}}
%		\item \newverold{In non-fully overlapping scenarios, cumulative interference and flow starvation may appear and cause poor performance to some WLANs.}
%		\item \newverold{There is not a unique optimal DCB policy.}
%	\end{enumerate}
	
	\subsection{Feasible states dependence on the DCB policy}
	
	% Explain DCB policy effect on scenario I. SCB and only-primary
	% In Scenario I there are $|\Psi| = 12$ global and $|\mathcal{S}| = 5$ feasible states when both WLANs apply \textit{always-max} DCB (see Figure \ref{tikz:scenarioI_ctmn}). However, depending on the policy, different states may be reachable and, therefore, the feasible state space and the average throughput experienced by each WLAN may vary accordingly.
	In Table \ref{table:cb_policy_effect} we show the effect of applying different DCB policies on the average throughput experienced by WLANs A and B ($\Gamma_\text{A}$ and $\Gamma_\text{B}$, respectively), and by the whole network ($\Gamma$) in scenarios \textit{I} and \textit{II} (presented in Figures \ref{fig:scenario_I} and \ref{fig:scenario_II}, respectively).
	% Explain CB policy effect on scenario I. DCB
	Let us first consider \textit{Scenario I}. As explained in Section \ref{sec:ctmn_model} and shown in Figure \ref{tikz:scenarioI_ctmn}, the CTMN reaches 5 feasible states when WLANs implement AM. Instead, due to the fact that both WLANs overlap in channels 3 and 4 when transmitting in their whole allocated channels -- i.e., $C_\text{A}^\text{tx} = C_\text{A} = \{1,2,3,4\}$ and $C_\text{B}^\text{tx} = C_\text{B} = \{3,4\}$, respectively -- the SCB policy reaches just three feasible states. Such states correspond to those with a single WLAN transmitting, i.e., $\mathcal{S} = \{\emptyset$, $\text{A}_1^4$, $\text{B}_3^4\}$. In the case of OP, both WLANs are forced to pick just their primary channel for transmitting and, therefore, $\mathcal{S} = \{\emptyset, \text{A}_2^2, \text{B}_3^3, \text{A}_2^2 \text{B}_3^3\}$. Notice that state $\text{A}_2^2 \text{B}_3^3$ is feasible because A and B have different primary channels and do not overlap when transmitting in them.
	
	\begin{table}[htbp]
		\footnotesize
		\centering
		\caption{\newverold{Tool features comparison.}}
		\label{table:sfctmn_vs_komondor}
		\begin{threeparttable}
			\begin{tabular}{@{}ccc@{}}
				\toprule
				Tool                  & SFCTMN & Komondor \\ \midrule
				Type	& Analytical model	& Simulator \\
				DCB policies          & \checkmark      & \checkmark              \\
				Hidden \& exposed nodes          & \checkmark      & \checkmark              \\
				RTS/CTS               & \xmark\tnote{(a)}      & \checkmark              \\
				Flow-in-the-middle    & \checkmark      & \checkmark              \\
				Information asymmetry & \checkmark      & \checkmark              \\
				Backoff collision & \xmark       & \checkmark              \\
				Scalable\tnote{(b)}              & \xmark      & \checkmark              \\ \bottomrule
			\end{tabular}
			\begin{tablenotes}
				\scriptsize \item $^\text{(a)}$RTS/CTS overhead considered in throughput measurements.
				\scriptsize \item $^\text{(b)}$Assumable execution time for up to 300 nodes.
				
			\end{tablenotes}
		\end{threeparttable}
		
	\end{table}

	\begin{table}[htbp]
		\small
		\centering
		\caption{DCB policy effect on the average throughput [Mbps] in \textit{Scenario I} and \textit{Scenario II}. The values obtained through \texttt{Komondor} are displayed in parentheses, while the other correspond to the SFCTMN framework.}
		\label{table:cb_policy_effect}
		\begin{tabularx}{.47\textwidth}{CCCCCCCCC}
			
			\toprule
			
			\multicolumn{1}{c}{\textbf{Policy}} & 
			\multicolumn{4}{c}{\textit{\textbf{Scenario I}}} & \multicolumn{4}{c}{\textit{\textbf{Scenario II}}} \\
			
			\multicolumn{1}{c}{$\mathcal{D}$}                    & $|\mathcal{S}|$              & $\Gamma_\text{A}$ & $\Gamma_\text{B}$ & $\overline{\Gamma}$ & $|\mathcal{S}|$             & $\Gamma_\text{A}$ & $\Gamma_\text{B}$ & $\overline{\Gamma}$ \\
			
			\midrule
			
			% OP
			\multicolumn{1}{c}{\multirow{2}{*}{OP}}  & \multirow{2}{*}{4}  & 109.36    & 109.36     & 109.36   & \multirow{2}{*}{4} & 109.36     & 109.36     & 109.36   \\ 
			\multicolumn{1}{c}{}                     &                     & \scriptsize (109.36)     & \scriptsize (109.36)     & \scriptsize (109.36)   &                    & \scriptsize (109.36)     & \scriptsize (109.36)     & \scriptsize (109.36)   \\ 
			
			% SCB
			\multicolumn{1}{c}{\multirow{2}{*}{SCB}} & \multirow{2}{*}{3}  & 132.75     & 132.75     & 132.75   & \multirow{2}{*}{3} & 102.65     & 102.65     & 102.65   \\
			\multicolumn{1}{c}{}                     &                     & \scriptsize (123.21)     & \scriptsize (137.09)     & \scriptsize (130.15)   &                    & \scriptsize (102.24)     & \scriptsize (102.24)     & \scriptsize (102.24)   \\ 
			
			% AM
			\multicolumn{1}{c}{\multirow{2}{*}{AM}}  & \multirow{2}{*}{5}  & 206.68    & 199.67    & 203.17   & \multirow{2}{*}{3} & 102.65     & 102.65     & 102.65   \\ 
			\multicolumn{1}{c}{}                     &                     & \scriptsize (204.70)     & \scriptsize (201.91)     & \scriptsize (203.31)   &                    & \scriptsize (102.24)     & \scriptsize (102.24)     & \scriptsize (102.24)   \\ 
			
			% PU
			\multicolumn{1}{c}{\multirow{2}{*}{PU}}  & \multirow{2}{*}{10} & 142.70    & 142.00     & 142.35   & \multirow{2}{*}{6} & 109.30     & 109.30     & 109.30   \\  
			\multicolumn{1}{c}{}                     &                     & \scriptsize (142.69)     & \scriptsize (142.01)     & \scriptsize (142.35)   &                    & \scriptsize (109.29)     & \scriptsize (109.27)     & \scriptsize (109.28)   \\ 
			\bottomrule
		\end{tabularx}
	\end{table}

	% Explain CB policy effect on scenario I. always-max
	% With AM the CTMN reaches 5 feasible states as shown in Figure \ref{tikz:scenarioI_ctmn}. Notice that with this policy, anytime the backoff of the AP in WLAN X expires and the system is in $\emptyset$ state, the AP will pick the widest available channel (i.e., $C_\text{X}^{\text{tx}} = C_\text{X}$) because all the basic channels in $C_\text{X}$ will be found idle as no packets are transmitted in state $\emptyset$. Therefore, the CTMN transits from $\emptyset$ to $\text{A}_1^4$ or to $\text{B}_3^4$ when the backoff of A or B expires, respectively. Nonetheless, whenever A's backoff counter terminates and the system is in state $B_3^4$, the widest allowed channel for transmitting is $C_\text{A}^\text{tx}=\{1,2\}$. Therefore, the CTMN will transit to state $\text{A}_1^2 \text{B}_3^4$, where both WLANs transmit concurrently. This is in fact the reason why state $\text{A}_1^2$ is reachable. Namely, when B finishes its transmission in state $\text{A}_1^2 \text{B}_3^4$, during the new backoff process, A keeps transmitting in the two channels of state $A_1^2$ until finishing. Therefore, even though forward transitions from $\emptyset$ to $A_1^2$ are not allowed by AM, a backward transition between such states is given when A finishes its transmission while B is still decreasing its backoff counter.
	
	% Explain CB policy effect on scenario I. Prob. uniform
	% Transition examples on PU. Alphas
	The last policy studied is PU, which is characterized by providing further exploration \newver{of} the global state space $\Psi$. It usually allows expanding the feasible state space $\mathcal{S}$ accordingly because more transitions are permitted. In \textit{Scenario I}, whenever the CTMN is in state $\emptyset$ and the backoff of A or B expires, the WLANs pick each of the possible available channels with \newver{the same} probability. Namely, the CTMN will transit to $\text{A}_2^2$, $\text{A}_1^2$ or $\text{A}_1^4$ with probability 1/3 when A's backoff counter terminates, and to $\text{B}_3^3$ or $\text{B}_3^4$ with probability 1/2 whenever B's backoff counter terminates.
	
	Likewise, if the system is in state $\text{B}_3^3$ and A terminates its backoff counter, the CTMN will transit to the feasible states $\text{A}_2^2 \text{B}_3^3$ or $\text{A}_1^2 \text{B}_3^3$ with probability 1/2. Similarly, whenever the system is in state $\text{A}_2^2$ or $\text{A}_1^2$, and B finishes its backoff, B will pick the transmission channels $\{3\}$ or $\{3,4\}$ with same probability 1/2 making the CTMN to transit to the corresponding state where both WLANs transmit concurrently. These probabilities are called transition probabilities and are represented by the vector $\vec{\alpha}_{\text{X},s}(s')$. For instance, in the latter case, the probability to transit from $s = \text{A}_2^2$ to $s' = \text{A}_2^2 \text{B}_3^3$ when B terminates its backoff is $\vec{\alpha}_{\text{B}, \text{A}_2^2}(\text{A}_2^2 \text{B}_3^3) = 1/2$.
	%\textcolor{cyan}{[Per el futur: En un sistema real, es fàcil estimar l'ocupació dels canals quan el node està en backoff o inactiu... i en funció d'això decidir els valors d'alfa.]} \sergio{[OK]}

	\subsection{\newverold{Instantaneous vs. long-term throughput}}
	
	% Use also Scenario I to talk about non magical policies
	
\begin{figure}[t]
		\centering
		\includegraphics[width=0.3\textwidth]{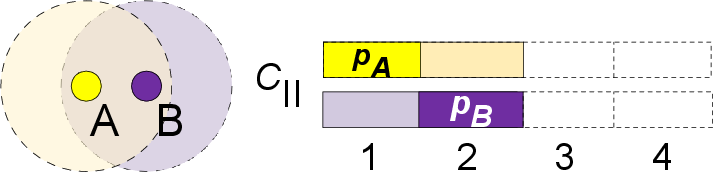}
		\caption{\textit{Scenario II}. WLANs A and B are inside the carrier sense range of each other with potentially overlapping basic channels 1 and 2.}
		\label{fig:scenario_II}
	\end{figure}

\begin{figure}
		\centering
		\begin{tikzpicture}[<->,>=stealth',shorten >=1pt,auto,node distance=1.4cm,
		semithick]
		\tikzstyle{every state}=[fill=white,draw=black,thick,text=black,scale=1]
		
		\node[state, label=below:$s_1$]    (S1b)                    {$\emptyset$};
		
		\node[state, label=below:$s_3$]    (S3b)[above right of=S1b, xshift=3.9cm]		{$A_{1}^{2}$};
		
		\node[state, label=below:$s_2$]    (S2b)[above of = S3b]		{$A_{1}^{1}$};
		
		\node[state, label=below:$s_4$]    (S4b)[below right of = S1b, xshift=3.9cm]		{$B_{2}^{2}$};
		
		\node[state, label=below:$s_5$]    (S5b)[below of= S4b]		{$B_{1}^{2}$};
		
		\node[state, label=below:$s_6$]    (S6b)[above right of =S4b, , xshift=2cm]		{$A_{1}^{1}B_2^2$};
		
		\path
		
		(S1b) edge[dashed, bend left]     node[sloped, anchor=center, above]{$\vec{\alpha}_{\text{A},\emptyset}(s_2)\lambda_\text{A},\mu_\text{A}(s_2)$} node[sloped, anchor=center, below]{\footnotesize \textcolor{blue}{$1,5$}} (S2b)
		
		(S1b) edge[dashed]     node[sloped, anchor=center, above]{$\vec{\alpha}_{\text{A},\emptyset}(s_3)\lambda_\text{A},\mu_\text{A}(s_3)$} 
		node[sloped, anchor=center, below]{\footnotesize \textcolor{blue}{$2,7$}} (S3b)
		
		(S1b) edge[dashed]     node[sloped, anchor=center, above]{$\vec{\alpha}_{\text{B},\emptyset}(s_4)\lambda_\text{B},\mu_\text{B}(s_4)$} 
		node[sloped, anchor=center, below]{\footnotesize \textcolor{blue}{$3,9$}}
		(S4b)
		
		(S1b) edge[dashed, bend right]     node[sloped, anchor=center, above] [yshift = 2]{$\vec{\alpha}_{\text{B},\emptyset}(s_5)\lambda_\text{B},\mu_\text{B}(s_5)$} 
		node[sloped, anchor=center, below]{\footnotesize \textcolor{blue}{$4,10$}} (S5b)
		
		(S2b) edge[dashed]   node[sloped, anchor=center, above]{$\lambda_\text{B},\mu_\text{B}(s_5)$} 
		node[sloped, anchor=center, below]{\footnotesize \textcolor{blue}{$6,12$}} (S6b)
		
		(S4b) edge[dashed]     node[sloped, anchor=center, above]{$\lambda_\text{A},\mu_\text{A}(s_5)$} 
		node[sloped, anchor=center, below]{\footnotesize \textcolor{blue}{$8,11$}} (S6b);
		
		\end{tikzpicture}
		\caption[]{CTMN corresponding to \textit{Scenario II}. Transitions edges are dashed for referring to those that may be given or not depending on the DCB policy. For instance, state $s_6$ is only reachable for the OP and PU policies. The discovery order of the states and transitions (displayed in blue) corresponds to the PU policy.}
		\label{tikz:scenarioII_ctmc}
	\end{figure}
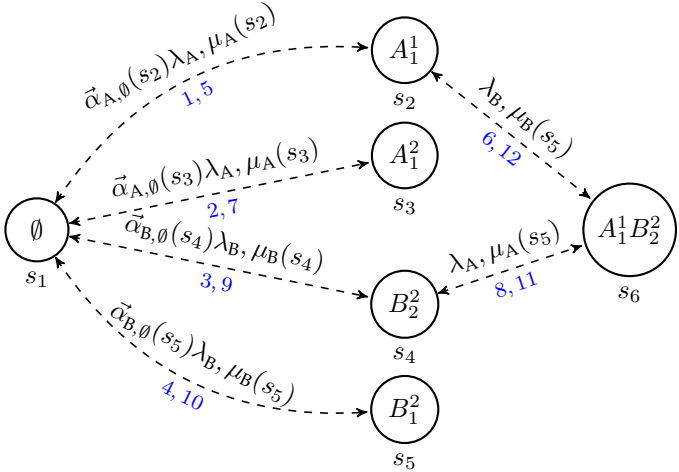

	% Intuition says that optimize short term throughput is better but that not happens in II
	Intuitively, one could think that, as it occurs in \textit{Scenario I}, always picking the widest channel found free by means of AM, i.e., maximizing the throughput of the immediate packet transmission (or instantaneous throughput), may be the best strategy for maximizing the long-term throughput as well. However, the \textit{Scenario II} depicted in Figure \ref{fig:scenario_II} is a counterexample that illustrates such lack of applicable intuition. It consists of two overlapping WLANs as in \textit{Scenario I}, but with different channel allocation: $C_\text{A}=C_\text{B}=\{1,2\}$ with $p_\text{A} = 1$ and $p_\text{B} =2$, respectively. The CTMNs that are generated according to the different DCB policies -- generalized to any value the corresponding transition probabilities $\vec{\alpha}_{\text{A},\emptyset}, \vec{\alpha}_{\text{B},\emptyset}$ may have -- are shown in Figure \ref{tikz:scenarioII_ctmc}.
	
	% Scenario II figure
	
	% Scenario II SFCTMN
	
	% ML learning or other techniques should be apply
	% Despite being a very simple scenario, it is not straightforward to determine the optimal DCB policy that the AP in each WLAN must follow to optimize its own long-term throughput or the system's aggregate. %Besides, it is important to emphasize that in these scenarios we are considering complete knowledge of the system and total control when generating the SFCTMN. Without such detail of information, designing a proper DCB policy would be even a harder task.
	
	% Explain alphas
	Regarding the transition probabilities, Table \ref{table:cb_policy_effect_alphas} shows the vectors $\vec{\alpha}_{\text{A},\emptyset}, \vec{\alpha}_{\text{B},\emptyset}$ that are given for each of the studied DCB policies in \textit{Scenario II}. Firstly, with OP, due to the fact that WLANs are only allowed to transmit in their primary channel, the CTMN can only transit from state $\emptyset$ to states $\text{A}_1^1$ or $\text{B}_2^2$, i.e., $\vec{\alpha}_{\text{A},\emptyset}(s_2)=\vec{\alpha}_{\text{B},\emptyset}(s_4) = 1$. Similarly, with SCB, WLANs can only transmit in their complete allocated channel, thus, when being in state $\emptyset$ the CTMN transits only to $\text{A}_1^2$ or $\text{B}_1^2$, i.e., $\vec{\alpha}_{\text{A},\emptyset}(s_3)=\vec{\alpha}_{\text{B},\emptyset}(s_5) = 1$. Notice that AM generates the same transition probabilities (and respective average throughput) than SCB because whenever the WLANs have the possibility to transmit -- which only happens when the CTMN is in state $\emptyset$ -- both A and B pick the widest channel available, i.e., $C_\text{A}^\text{tx} = C_\text{B}^\text{tx} = \{1,2\}$. Finally, PU picks uniformly at random any of the possible transitions that A and B provoke when terminating their backoff in state $\emptyset$, i.e., $\vec{\alpha}_{\text{A},\emptyset}(s_2)=\vec{\alpha}_{\text{A},\emptyset}(s_3) = 1/2$ and $\vec{\alpha}_{\text{B},\emptyset}(s_4)=\vec{\alpha}_{\text{B},\emptyset}(s_5) = 1/2$, respectively.
	
	% How s6 is reached.
	Note that state $\text{A}_1^1 \text{B}_2^2$, when both WLANs are transmitting at the same time, is reachable from states $\text{A}_1^1$ and $\text{B}_2^2$ for both OP and PU. In such states, when either A or B terminates its backoff and the other is still transmitting in its primary channel, only a transition to state $\text{A}_1^1 \text{B}_2^2$ is possible, i.e.,  $\vec{\alpha}_{\text{A},s_2}(s_6)=\vec{\alpha}_{\text{B},s_4}(s_6) = 1$.
	
	% Transition rate probabilities table
	%	\begin{table}[h]
	%		\centering
	%		\caption{Transition probabilities from state $\emptyset$ of WLANs A and B in \textit{Scenario II} for different DCB policies.}
	%		\label{table:cb_policy_effect_alphas}
	%		\normalsize
	%		\begin{tabular}{|c|c|c|c|c|c|}
	%			\hline
	%			$\mathcal{D}$ & $|\mathcal{S}|$ & $\vec{\alpha}_{\text{A},\emptyset}(s_2)$ & $\vec{\alpha}_{\text{A},\emptyset}(s_3)$ & $\vec{\alpha}_{\text{B},\emptyset}(s_4)$ & $\vec{\alpha}_{\text{B},\emptyset}(s_5)$ \\ \hline
	%			OP            & 4               & 1.0             & 0.0             & 1.0             & 0.0             \\ \hline
	%			SCB           & 3               & 0.0             & 1.0             & 0.0             & 1.0             \\ \hline
	%			AM            & 3               & 0.0             & 1.0             & 0.0             & 1.0             \\ \hline
	%			PU            & 6               & 0.5             & 0.5             & 0.5             & 0.5             \\ \hline
	%		\end{tabular}
	%		
	%	\end{table}
	
	\begin{table}[htbp]
		\small
		\centering
		\caption{Transition probabilities from state $\emptyset$ of WLANs A and B in \textit{Scenario II} for different DCB policies.}
		\label{table:cb_policy_effect_alphas}
		\begin{tabularx}{.47\textwidth}{CCCCCC}
			\toprule
			
			$\boldsymbol{\mathcal{D}}$ & 
			$\boldsymbol{|\mathcal{S}|}$ & 
			$\boldsymbol{\vec{\alpha}_{\text{A},\emptyset}(s_2)}$ &
			$\boldsymbol{\vec{\alpha}_{\text{A},\emptyset}(s_3)}$ & $\boldsymbol{\vec{\alpha}_{\text{B},\emptyset}(s_4)}$ & $\boldsymbol{\vec{\alpha}_{\text{B},\emptyset}(s_5)}$ \\
			
			\midrule
			OP            & 4               & 1.0             & 0.0             & 1.0             & 0.0             \\ 
			SCB           & 3               & 0.0             & 1.0             & 0.0             & 1.0             \\ 
			AM            & 3               & 0.0             & 1.0             & 0.0             & 1.0             \\ 
			PU            & 6               & 0.5             & 0.5             & 0.5             & 0.5             \\ 
			\bottomrule
		\end{tabularx}
	\end{table}

	% Scenario III
	\begin{figure*}[t!]
		\centering
		\begin{subfigure}{.46\textwidth}
			\centering
			\includegraphics[width=.77\linewidth]{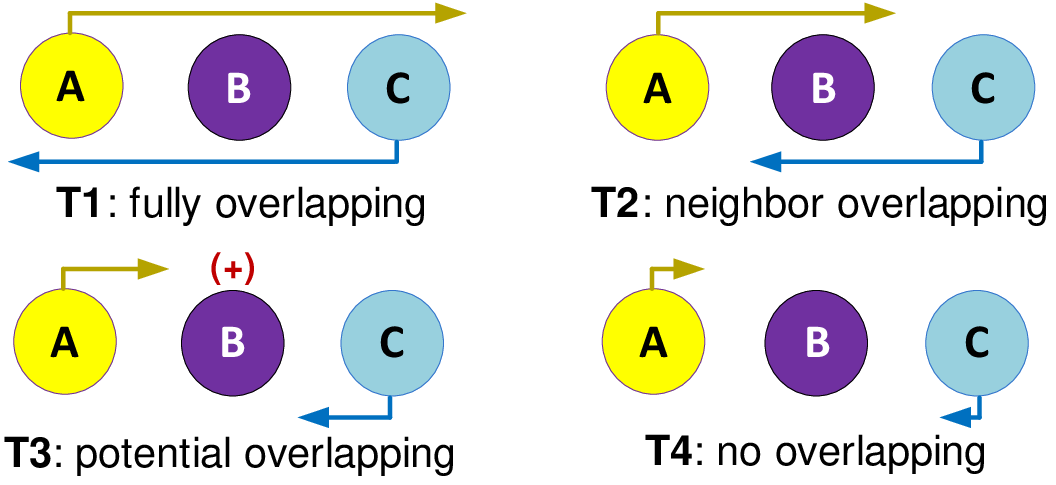}
			\caption{Topologies.}
			\label{fig:scenario_III}
		\end{subfigure}%
		\begin{subfigure}{.48\textwidth}
			\centering
			\includegraphics[width=0.99\linewidth]{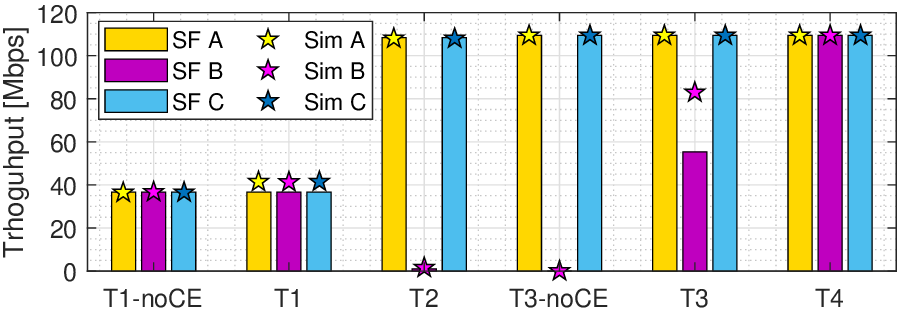}
			\caption{Average throughput.}
			\label{fig:scenario3_results_sfctmn_komondor}
		\end{subfigure}
		\caption{\textit{Scenario III}. Yellow and blue arrows indicate the carrier sense range of WLANs A and C, respectively. \textit{T1-noCE} and \textit{T3-noCE} refer to topologies \textit{T1} and \textit{T3} when B does not accomplish the capture effect condition. \textit{SF} and \textit{Sim} refer to the values obtained through SFCTMN and \texttt{Komondor}, respectively.}
		\label{fig:scenario_III_all}
	\end{figure*}

	% Interesting that OP performs the best.
	Interestingly, as shown in Table \ref{table:cb_policy_effect}, applying OP in\textit{ Scenario II}, i.e., being conservative and unselfish, is the best policy to increase both the individual average throughput of A and B ($\Gamma_\text{A}$, $\Gamma_\text{B}$, respectively) and the system's aggregated one ($\Gamma$). Instead, being aggressive and selfish, i.e., applying SCB or AM, provides the worst results both in terms of individual and system's aggregate throughput.
	% Why PU and OP perform similarly
	In addition, PU provides similar results than OP in average because most of the times that A and B terminate their backoff counter, they can only transmit in their primary channel since the secondary channel is most likely occupied by the other WLAN. In fact, state $\text{A}_1^1 \text{B}_2^2$ is the dominant state for both OP and PU. Specifically, the probability of finding the CTMN in state $\text{A}_1^1 \text{B}_2^2$, i.e., $\vec{\pi}_{s_6}$, is 0.9802 for OP and 0.9702 for PU, respectively. Therefore, the slight differences on throughput experienced with OP and PU are given because of the possible transition from $\emptyset$ to the states $A_1^2$ and $B_1^2$ in PU, where WLANs entirely occupy the allocated channel, thus preventing the other for decreasing its backoff.
	
	% Conclusion: some intelligence is required
	Despite being a very simple scenario, we have shown that it is not straightforward to determine the optimal DCB policy that the AP in each WLAN must follow. Evidently, in a non-overlapping scenario, AM would be the optimal policy for both WLANs because of the non-existence of inter-WLAN contention. However, this is not a typical case in dense scenarios and, consequently, AM may not be adopted as de facto DCB policy, even though it provides more flexibility than SCB. This simple toy scenario also serves to prove that some intelligence should be implemented in the APs in order to harness the information gathered from the environment.
	
	% Differences between SFCTM and Komondor I
	Concerning the throughput differences in the values obtained by SFCTMN and \texttt{Komondor},\footnote{In \texttt{Komondor} the throughput is simply computed as the number of useful bits (corresponding to data packets) that are successfully transmitted divided by the observation time of the simulation.} we note that the main disparities correspond to the AM and SCB policies. It is important to remark that while SFCTMN \newver{considers neither} backoff collisions nor NAV periods, \texttt{Komondor} actually does so in a more realistic way. Therefore, in \texttt{Komondor}, whenever there is a slotted backoff collision, the RTS packets can be decoded by the STAs in both WLANs if the CE is accomplished. That is why the average throughput is consequently increased.
	
	% Differences between SFCTM and Komondor II
	Regarding the NAV periods, an interesting \newver{phenomenon occurs} in \textit{Scenario I} when implementing SCB, AM or PU. While the RTS packets sent by B cannot be decoded by A because its primary channel is always outside the possible transmission channels of B (i.e., $p_\text{A}=2 \notin C_\text{B}^\text{tx}=\{3,3\}$ or $\{3,4\}$), the opposite occurs when A transmits them. Due to the fact that the RTS is duplicated in each of the basic channels used for transmitting, whenever A transmits in its whole allocated channel, B is able to decode the RTS (i.e., $p_\text{B}=3 \in C_\text{A}^\text{tx} = \{1,2,3,4\}$) and enters in NAV consequently.
	
	\subsection{\newverold{Cumulative interference and flow starvation}}
	
	% Introduce non fully overlapping scenarios and scenario I.
	When considering non-fully overlapping scenarios, i.e., where some of the WLANs are not inside the carrier sense range of the others, complex and hard to prevent phenomena may occur. As an illustrative example, let us consider the case shown in Figure \ref{fig:scenario_III}, where 3 WLANs sharing a single channel (i.e., $C_\text{A} = C_\text{B} = C_\text{C} = \{1\}$) are deployed composing a line network. 
	As the carrier sense range is fixed and is the same for each AP, by locating the APs at different distances we obtain different topologies that are worth to be analyzed. We name these topologies from \textit{T1} to \textit{T4} depending on the distance between consecutive APs, which increases according to the topology index. Notice that all the DCB policies discussed in this work behave exactly the same way in single-channel scenarios. Therefore, in this subsection, we do not make distinctions among them.
	
	% Results full overlapping T1
	The average throughput experienced by each WLAN in each of the regions is shown in Figure \ref{fig:scenario3_results_sfctmn_komondor}. Regarding topology \textit{T1}, when APs are close enough to be inside the carrier sense range of each other in a fully overlapping manner, the medium access is shared fairly because of the CSMA/CA mechanism. For that reason, the throughput is decreased to approximately 1/3 with respect to topology \textit{T4}. Specifically, the system spends almost the same amount of time in the states where just one WLAN is transmitting, i.e., $\pi(\text{A}_1^1) = \pi(\text{B}_1^1) = \pi(\text{C}_1^1) \approx 1/3$.
	% Results neighbor overlapping T2
	The neighbor overlapping case in topology \textit{T2} \newver{is a clear case of flow-in-the-middle (FIM) starvation. Note that A and C can transmit at the same time whenever B is not active, but B can only do so when neither A nor C are active.} Namely, B has very few transmission opportunities because A and C are transmitting almost permanently and B must continuously pause its backoff consequently.
	
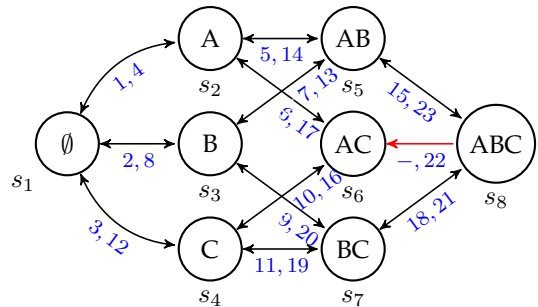
\begin{figure}[t]
		\centering
		\begin{tikzpicture}[<->,>=stealth',shorten >=1pt,auto,node distance=1.4cm,
		semithick]
		\tikzstyle{every state}=[fill=white,draw=black,thick,text=black,scale=1]
		
		\node[state, label=below left:$s_1$]    (S1)                    {$\emptyset$};
		
		\node[state, label=below:$s_3$]    (S3)[right of = S1, xshift=0.5cm]		{$\text{B}$};
		
		\node[state, label=below:$s_2$]    (S2)[above of = S3]		{$\text{A}$};
		
		\node[state, label=below:$s_4$]    (S4)[below of = S3]		{$\text{C}$};
		
		\node[state, label=below:$s_6$]    (S6)[right of = S3, xshift=0.5cm]		{$\text{A}\text{C}$};
		
		\node[state, label=below:$s_5$]    (S5)[above of = S6]		{$\text{A}\text{B}$};
		
		\node[state, label=below:$s_7$]    (S7)[below of = S6]		{$\text{B}\text{C}$};
		
		\node[state, label=below:$s_8$]    (S8)[right of = S6, xshift=0.5cm]		{$\text{A}\text{B}\text{C}$};
		
		\path
		
		(S1) edge[bend left]  node[sloped, anchor=center, above]{} node[sloped, anchor=center, below]{\footnotesize \textcolor{blue}{$1,4$}} (S2)
		
		(S1) edge    node[sloped, anchor=center, above]{} node[sloped, anchor=center, below]{\footnotesize \textcolor{blue}{$2,8$}} (S3)
		
		(S1) edge[bend right]    node[sloped, anchor=center, above]{} node[sloped, anchor=center, below]{\footnotesize \textcolor{blue}{$3,12$}} (S4)
		
		(S2) edge    node[sloped, anchor=center, above]{} node[sloped, anchor=center, below]{\footnotesize \textcolor{blue}{$5,14$}} (S5)
		
		(S2) edge    node[sloped, anchor=center, above]{} node[sloped, anchor=center, below]{\footnotesize \textcolor{white}{empty} \textcolor{blue}{$6,17$}} (S6)
		
		(S3) edge    node[sloped, anchor=center, above]{} node[sloped, anchor=center, below]{\footnotesize \textcolor{white}{empty} \textcolor{blue}{$7,13$}} (S5)
		
		(S3) edge    node[sloped, anchor=center, above]{} node[sloped, anchor=center, below]{\footnotesize \textcolor{white}{empty} \textcolor{blue}{$9,20$}} (S7)
		
		(S4) edge    node[sloped, anchor=center, above]{} node[sloped, anchor=center, below]{\footnotesize \textcolor{white}{empty} \textcolor{blue}{$10,16$}} (S6)
		
		(S4) edge    node[sloped, anchor=center, above]{} node[sloped, anchor=center, below]{\footnotesize \textcolor{blue}{$11,19$}} (S7)
		
		(S5) edge    node[sloped, anchor=center, above]{} node[sloped, anchor=center, below]{\footnotesize \textcolor{blue}{$15,23$}} (S8)
		
		(S6) edge[<-, red]    node[sloped, anchor=center, above]{} node[sloped, anchor=center, below]{\footnotesize \textcolor{blue}{$-,22$}} (S8)
		
		(S7) edge    node[sloped, anchor=center, above]{} node[sloped, anchor=center, below]{\footnotesize \textcolor{blue}{$18,21$}} (S8);

		\end{tikzpicture}
		\caption[]{CTMN corresponding to \textit{Scenario III-T3}. For the sake of visualization, neither the transition rates nor the transmission channels are included in the figure. The discovery order of the transitions is represented by the pairs in blue.}
		\label{tikz:scenario3_ctmn}
	\end{figure}

	% Results potential central node overlapping T3
	An interesting and hard to prevent \newver{phenomenon} occurs in the potential central node overlapping case at topology \textit{T3}. Figure \ref{tikz:scenario3_ctmn} shows the corresponding CTMN. \newver{In this case, the cumulated interference perceived by B from both A and C, prevents the former to decrease its backoff, thus generating a new FIM-like scenario such as in topology \textit{T2}}. However, in this case, B is able to decrement the backoff any time A or C are not transmitting.
	% Potential overlaping explanation
	This leads to two possible outcomes regarding packet collisions. On the one hand, if the capture effect condition is accomplished by B (i.e., $\gamma_\text{B} > \text{CE}$) no matter whether A and C are transmitting, B will be able to successfully exchange packets and the throughput will increase accordingly. On the other hand, if the capture effect condition is not accomplished, B will suffer a huge packet error rate because most of the initiated transmissions will be lost due the hidden node effect caused by the concurrent transmissions of A and C (i.e., $\gamma_\text{B} < \text{CE}$ when A and C transmit). This phenomena may be recurrent and have considerable impact \newver{on} \newver{high-density} networks where multiples WLANs interact with each other. Therefore, it should be foreseen in order to design efficient DCB policies.
	
	% Results isolation T4
	\newver{Finally, in topology \textit{T4}, WLANs achieve the maximum throughput, as expected. The fact is that they are isolated (i.e., outside the carrier sense range of each other), which allows holding successful transmissions without having to pause their backoff.}
	
	% SFCTMN scenario III region 3

	% Why throughput differences between SFCTMN and Komondor
	Concerning the differences on the average throughput values estimated by SFCTMN and \texttt{Komondor}, we observe two phenomena with respect to backoff collisions in topologies \textit{T1} and \textit{T3}. In \textit{T1}, due to the fact that simultaneous transmissions (or backoff collisions) are permitted and captured in \texttt{Komondor}, the throughput is slightly smaller or higher depending on whether the capture effect condition is accomplished (T1) or not (T1-noCE), respectively. Note that backoff collisions have a negligible effect in T2 since B suffers from heavy FIM and it hardly ever transmits.
	
	% T3 Non-insensitivity
	The most notable difference is given in \textit{T3}. In this topology, SFCTMN estimates that B is transmitting just the 50.15\% of the time. That is, since A and C operate like in isolation, most of the time they transmit concurrently, causing backoff freezing at B. However, \texttt{Komondor} estimates that B transmits about the 75\% of the time, capturing a more realistic behavior. Such a difference is caused by the fact that the insensitivity property does not hold in this setup, since the Markov chain is not reversible. For instance, whenever the system is in state $s_6 = \text{A}_1^1 \text{C}_1^1$ and A finishes its transmission (transiting to $s_4 = C_1^1)$, B decreases its backoff accordingly while C is still active. Therefore, it is more probable to transit from $s_4$ to $s_7 = \text{B}_1^1 \text{C}_1^1$ than to $s_6 = \text{A}_1^1 \text{C}_1^1$ again because, in average, the remaining backoff counter of B will be smaller than the generated by A when finishing its transmission. This is in fact not considered by the CTMN, which assumes the same probability to transit from $s_4$ to $s_6$ than to $s_7$ because of the exponential distribution and the memoryless property.

	%-------------------------
	\subsection{Variability of optimal policies}
	
	\begin{table}[htbp]
		\footnotesize
		\centering
		\caption{\newver{Policy combinations effect on throughput and fairness in the WLANs of \textit{Scenario IV}. The complete table can be found at Appendix B of the supplementary material.}}
		\label{table:scenario_IV}
		\begin{tabularx}{.48\textwidth}{vvvCCCCCC}
			\toprule
			\multicolumn{3}{c}{\textbf{Policy}}	& \textbf{States}
			& \multicolumn{4}{c}{\textbf{Throughput [Mbps]}}	& \textbf{Fair.}		\\
			
			$\mathcal{D}_A$ & $\mathcal{D}_B$ & $\mathcal{D}_C$ & $\abs S$ & $\Gamma_\text{A}$  & $\Gamma_\text{B}$  & $\Gamma_\text{C}$  & $\Gamma$   & $\mathcal{J}$ \\
			
			\midrule
			
			% AM AM AM
			AM & AM  & AM  & 5  & 199.96     & 3.58   & 199.96 & \textbf{403.49}     &  0.67853   \\
			
			% AM PU AM
			AM & PU  & AM  & 10 & 149.41     & 62.45   & 149.41 & 361.27     & 0.89679   \\
			
			% PU AM PU
			PU & AM  & PU   & 25 & 109.84     & 108.44   & 109.84 & 328.12     & \textbf{0.9999}   \\
			
			% AM AM PU
			AM & AM  & PU   & 9 & 111.31     & 106.91   & 110.33 & 328.55         & \textbf{0.9997}   \\
			%
			%			\multicolumn{1}{c}{} & & & & \scriptsize (109.64)     & \scriptsize (109.06)     &  (109.49)	   &   \scriptsize (328.19)     & \scriptsize (1.00000)   \\ 
			%			
			%			% AM PU PU
			%			\multirow{2}{*}{AM}  & \multirow{2}{*}{PU}  & \multirow{2}{*}{PU}  & \multirow{2}{*}{12} & 111.29     & 106.94   & 110.33     & 328.56     &  \textbf{0.99971}   \\
			%			
			%			\multicolumn{1}{c}{} & & & & \scriptsize (109.63)     & \scriptsize (109.07)     & \scriptsize (109.49)	   &   \scriptsize (328.18)     & \scriptsize (1.00000)   \\ 
			%			
			%			% PU PU PU
			%			\multirow{2}{*}{PU}  & \multirow{2}{*}{PU}  & \multirow{2}{*}{PU}  & \multirow{2}{*}{14} & 109.85     & 108.44   & 109.85 & 328.13     & \textbf{0.99996}  \\
			%			
			%			\multicolumn{1}{c}{} & & & & \scriptsize (109.52)     & \scriptsize (109.10)     & \scriptsize (109.51)	   &   \scriptsize (328.13)     & \scriptsize (1.00000)   \\ 
			
			\bottomrule
		\end{tabularx}
	\end{table}

	% Normally AM to increase own throughpu. But exceptions
	Most often, the best DCB policy for increasing the own throughput, no matter what policies the rest of WLANs may implement, is AM. Nonetheless, there are exceptions like the one presented in \textit{Scenario II}. Besides, if achieving throughput fairness between all WLANs is the objective, other policies may be required. Therefore, there is not always an optimal common policy to be implemented by all the WLANs. In fact, there are cases where different policies must be assigned to different WLANs in order to increase both the fairness and individual throughput. 
	% Aggregated vs. fairness with AM
	For instance, let us consider another toy scenario (\textit{Scenario IV}) using the topology \textit{T2} of \textit{Scenario III}, where three WLANs are located in a line in such a way that they are in the carrier sense range of the immediate neighbor. In this case, however, let us assume a different channel allocation: $C_\text{A} = C_\text{B} = C_\text{C} = \{1,2\}$ and $p_\text{A} = p_\text{C} = 1$, $p_\text{B} = 2$.
	
	Table \ref{table:scenario_IV} shows the individual and aggregated throughputs, and the Jain's fairness index for different combinations of DCB policies. We note that, while implementing AM in all the WLANs the system's aggregated throughput is the highest (i.e., $\Gamma = 403.49$ Mbps), the throughput experienced by B is the lowest (i.e., $\Gamma_\text{B} = 3.58$ Mbps), leading to a very unfair FIM situation as indicated by $\mathcal{J} \approx 0.69$.
	% AM is not always the best even for own throughput
	We also find a case where implementing AM does not maximize the individual throughput of B. Namely, when A and C implement AM (i.e., $\mathcal{D}_A = \mathcal{D}_C = \text{AM}$), it is preferable for B to implement PU and force states in which A and C transmit only in their primary channels. This increases considerably both the throughput of B and the fairness accordingly.

	% Most fair combinations
	Looking at the \newver{fairest} combinations, we notice that A, C or both must implement PU in order to let B transmit with \newver{a} similar amount of opportunities. This is achieved by the stochastic nature of PU, which lets the CTMN to explore more states. Accordingly, B experiences the highest throughput and the system achieves complete fairness (i.e., $\mathcal{J} \approx 1$). Nonetheless, the price to pay is to significantly decrease the throughput of A and C. In this regard, other fairness metrics like $\mathcal{P}$ would determine the optimality of a certain \newver{combination} of policies in a different way. For instance, in the scenario under evaluation, the combination providing the highest proportional fairness is AM-PU-AM (i.e., $\mathcal{P} \approx 6.16$) followed closely by the rest of scenarios with some WLAN implementing PU (i.e., $\mathcal{P} \approx 6.11$).
	
	In essence, this toy scenario is a paradigmatic example showing that with less aggressive policies like PU (of probabilistic nature), not only more states in the CTMN can be potentially explored with respect to AM, but also the probability of staying in states providing higher throughput (or fairness) may increase. Therefore, since most of the times it does not exist a global policy that satisfies all the WLANs in the system, different policies should be adopted depending on the parameter to be optimized.
	
	%-------------------------
	
	%	\subsection{Time between states}
	%	
	%	\textcolor{red}{DISCUSS WITH BORIS}
	%	The transition time between dominant states (interesting to have a graph about it!)
	%	The number of packets lost and average throughput taking into account such losses. A packet is transmitted when a backward transition is done.
	%	Once you have the Markov network, it is possible to simulate it and determine:

	%-------------------------
	%-------------------------
	%-------------------------
	%-------------------------
	
	\section{Evaluation of DCB in \newver{dense} WLANs}	\label{sec:evaluation}
	
	% Intro to the section
	\newver{In this section, we study the effects of the presented DCB policies by simulating WLAN deployments of different node densities in \texttt{Komondor}. We first draw some general conclusions from analyzing the throughput and fairness when increasing the number of WLANs per area unit. Then, we discuss what is the optimal policy that a particular WLAN should locally pick in order to maximize its own throughput.
	% Evaluation setup in appendix
	The evaluation setup (11ax parameters, transmission power, path loss model, etc.) is extensively detailed in Appendix C of the supplementary material.}
	
	%\newverold{The results gathered in this section have been obtained using \texttt{11axHDWLANSim}, an event-based wireless network simulator that aims to capture the effects of the newest technologies included in the IEEE 802.11 amendments. This simulator is a particular release (v1.0.1) of Komondor~\cite{barrachina2018komondor},\footnote{\texttt{11axHDWLANSim} validation report v0.1: \url{https://github.com/wn-upf/Komondor/blob/master/Documentation/Other/validation_report_v01.pdf}} a wireless networks simulator built on top of the COST library \cite{chen2002reusing}.} \newver{The evaluation setup is extensively detailed in Appendix A.}
	
	\subsection{Network density vs. throughput}
	
	% Describe scenario generally
	Figure \ref{fig:map_central} shows the general scenario considered for conducting the experiments presented in this section. For the sake of speeding up \texttt{Komondor} simulations, we assume that each WLAN is composed just of 1 AP and 1 STA. Due to the fact that APs and STAs are located randomly in the map, the number of STAs should not have a significant impact on the results because only downlink single-user traffic is assumed.
	%Specific details
	Essentially, we consider a rectangular area $A_\text{map} = 100 \times 100$ m$^2$, where $M$ WLANs are placed uniformly at random with the single condition that any pair of APs must be separated at least $d_{\text{AP-AP}}^\text{min} = 10$ m. The STA of each WLAN is located also uniformly at random at a distance $d_\text{AP-STA} \in [d_\text{AP-STA}^\text{min}, d_\text{AP-STA}^\text{max}] = [1, 5]$ m from the AP. The channelization $\mathcal{C}$ counts with $N_\text{sys} = 8$ basic channels (160 MHz) and follows the 11ax proposal (see Figure \ref{fig:channel_allocation_ac}). Channel allocation is also set uniformly at random, i.e., every WLAN $w$ is assigned a primary channel $p_\text{w} \sim U[1,8]$ and allocated channel $C_\text{X}$ containing $N_\text{w} \sim U\{1,2,4,8\}$ basic channels.

	% How is the process of expermients
	For each number of WLANs studied (i.e., $M = 2,5,10,20,30,40,50$), we generate $N_\text{D} = 50$ deployments with different random node locations and channel allocations. Then, for each of the deployments, we assign to all the WLANs the same DCB policy. Namely, we simulate $N_\text{M} \times N_\text{D} \times N_\text{P} = 7 \times 50 \times 4 = 1400$ scenarios, where $N_\text{M}$ is the number of different $M$ values studied and $N_\text{P}$ is the number of DCB policies considered in this paper. Besides, all the simulations have a duration of $T_\text{obs} = 20$ seconds.
	
\begin{figure}[t]
		\centering
		\includegraphics[width=0.45\textwidth]{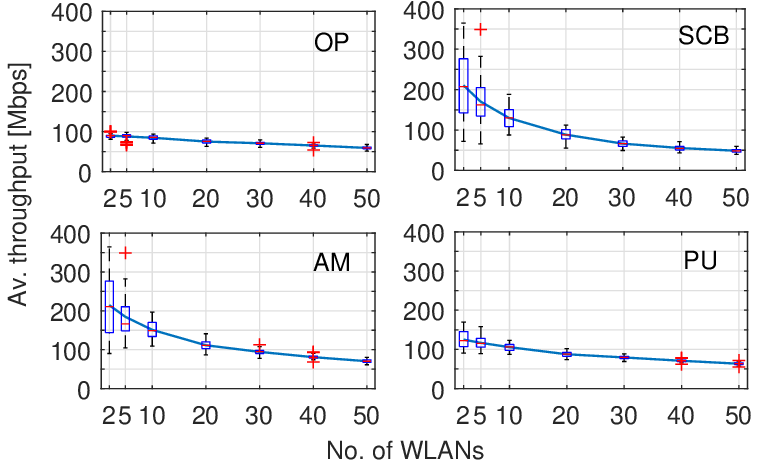}
		\caption{Node density effect on the average WLAN throughput. On each box, the central mark indicates the median, and the bottom and top edges of the box indicate the 25th and 75th percentiles, respectively. The whiskers extend to the most extreme data points not considered outliers, and the outliers are plotted individually using the `\Plus' symbol.}      
		\label{fig:results_av_throughput_per_policy_boxplot}
	\end{figure}

	% Discuss throughput.
	\newver{In Figure \ref{fig:results_av_throughput_per_policy_boxplot}, we show by means of boxplots the average throughput per WLAN for each of the presented DCB policies. As expected, when there are few WLANs in the area, the most aggressive policies (i.e., SCB and AM) provide the highest throughput. \newver{In contrast}, PU, and especially OP, perform the worst, as they do not extensively exploit the free bandwidth.}
	However, when the scenario gets denser, the average throughput obtained by all the policies except SCB tends to be similar. \newver{This occurs because WLANs implementing AM or PU tend to carry out single-channel transmissions since the PIFS condition for multiple channels are most likely not accomplished}. In the case of SCB, \newver{part of the bandwidth} in the WLAN's allocated channel will most likely be occupied by other WLANs, and therefore its backoff counter will get repeatedly paused. Thus, its average throughput in dense scenarios is considerably low with respect to the other policies.
	
\begin{figure}[t]
		\centering
		\includegraphics[width=0.48\textwidth]{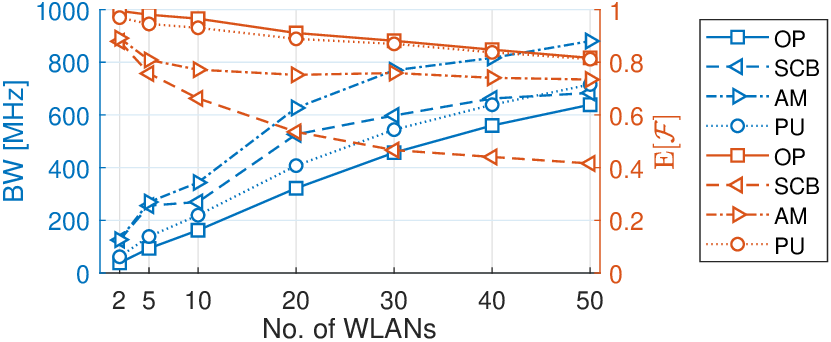}
		\caption{Node density effect on total bandwidth and fairness.}
		\label{fig:fairness_and_bw}
	\end{figure}

	% Spectrum utilization
	In order to asses the use of the spectrum, we first define the average bandwidth usage of a WLAN $w$ as $\text{E}[\text{BW}_w] = \frac{1}{T_\text{obs}}\sum_{c=1}^{N_\text{sys}} t_w^\text{tx}(c) \cdot |c|$, where $t_w^\text{tx}(c)$ is the time that WLAN $w$ is transmitting in a channel containing at least the basic channel $c$. The average spectrum used by all the WLANs, i.e., $\text{BW} = \sum_{w=1}^{M} \text{E}[\text{BW}_w]$ is shown in Figure \ref{fig:fairness_and_bw} for the different DCB policies.
	% Explain spectrum figure
	Similarly to the throughput, while OP and PU do not leverage the free spectrum in \newver{low-density} scenarios, SCB and AM do so by exploiting the most bandwidth. Instead, when the number of nodes per area increases, SCB suffers from heavy contention periods, which reiterates the need \newver{for} flexibility to adapt to the channel state. In this regard, we note that AM is clearly the policy exploiting the most bandwidth in average for any number of WLANs.
	% Fairness
	Nonetheless, neither the average throughput per WLAN nor the spectrum utilization may be a proper metric when assessing the performance of the whole system. Namely, having some WLANs experiencing high throughput when some others starve is often a situation preferable to be avoided. In that sense, we focus on the fairness, which is both indicated by the \textit{boxes} and outliers in Figure \ref{fig:results_av_throughput_per_policy_boxplot}, and more clearly represented by the expected Jain's fairness index shown in Figure \ref{fig:fairness_and_bw}.
	
	% Figure Throughput density
	
	% Figure spectrum utilization

	% OP and PU
	As expected, the policy providing the highest fairness is OP. In fact, no matter the channel allocation, WLANs only pick their primary channel for transmitting when implementing OP; hence the fairness is always maximized at the cost of probably wasting part of the frequency spectrum, \newver{especially} when the node density is low. In this regard, PU also provides high fairness while exploiting the spectrum more, which increases the average throughput per WLAN accordingly.
	% Comment on AM and SCB and boxes.
	Regarding the aggressive policies, SCB is clearly the most unfair policy due to its \textit{all or nothing} strategy. Therefore, it seems preferable to prevent WLANs from applying SCB in dense scenarios because of the number of WLANs that may starve or experience really low throughput. However, even though being aggressive, AM is able to adapt its transmission channel to the state of the medium, thus providing both higher throughput and fairness.
	% AM not always good
	Still, as indicated by the \textit{boxes} and \textit{outliers} of Figure \ref{fig:results_av_throughput_per_policy_boxplot}, AM is not \textit{per se} the optimal policy. In fact, there are scenarios where PU performs better in terms of both fairness and throughput. Consequently, there is room to improve the presented policies with some smarter adaptation or learning approaches (e.g., tunning properly the transition probabilities $\vec{\alpha}$ when implementing stochastic DCB).
	
	% Some interesting phenomena: BO slowness
	There are also some phenomena that we have observed during the simulations that are worth to be mentioned. Regarding backoff decreasing slowness, it can be the case that a WLAN $w$ is forced to decrease its backoff counter very slowly due to the fact that neighboring WLANs operate in a channel including the primary channel of $w$. That is why more fairness is achieved with PU in dense networks as such neighboring WLANs do not always pick the whole allocated channel. Thus, they let $w$ to decrease their backoff more often, and to proceed to transmit accordingly.
	% Power and channel range
	Finally, concerning the transmission power and channel width, we have observed that transmitting just in the primary channel can also be harmful to other WLANs because of the higher transmission power used per 20 MHz channel. While this may allow using higher MCS and respective data rates, it may also cause packet losses in neighboring WLANs operating with the same primary channel due to heavy interference, \newver{especially} for OP.
	
	\subsection{Local optimal policy}
	
	%\sergio{\textbf{Reviewer 3}: My concern is that the optimal policy may vary based on the density of the nodes. Particularly, the number of hidden nodes in the deployment scenario may impact on the choice of the optimal policy. Hence, I would like to suggest the authors to consider three deployment scenarios i.e., dense, semi-dense and sparse. In each scenario, the authors can investigate the optimal policy for channel bonding and make conclusions}
	
\begin{figure}[t]
		\centering
		\includegraphics[width=0.42\textwidth]{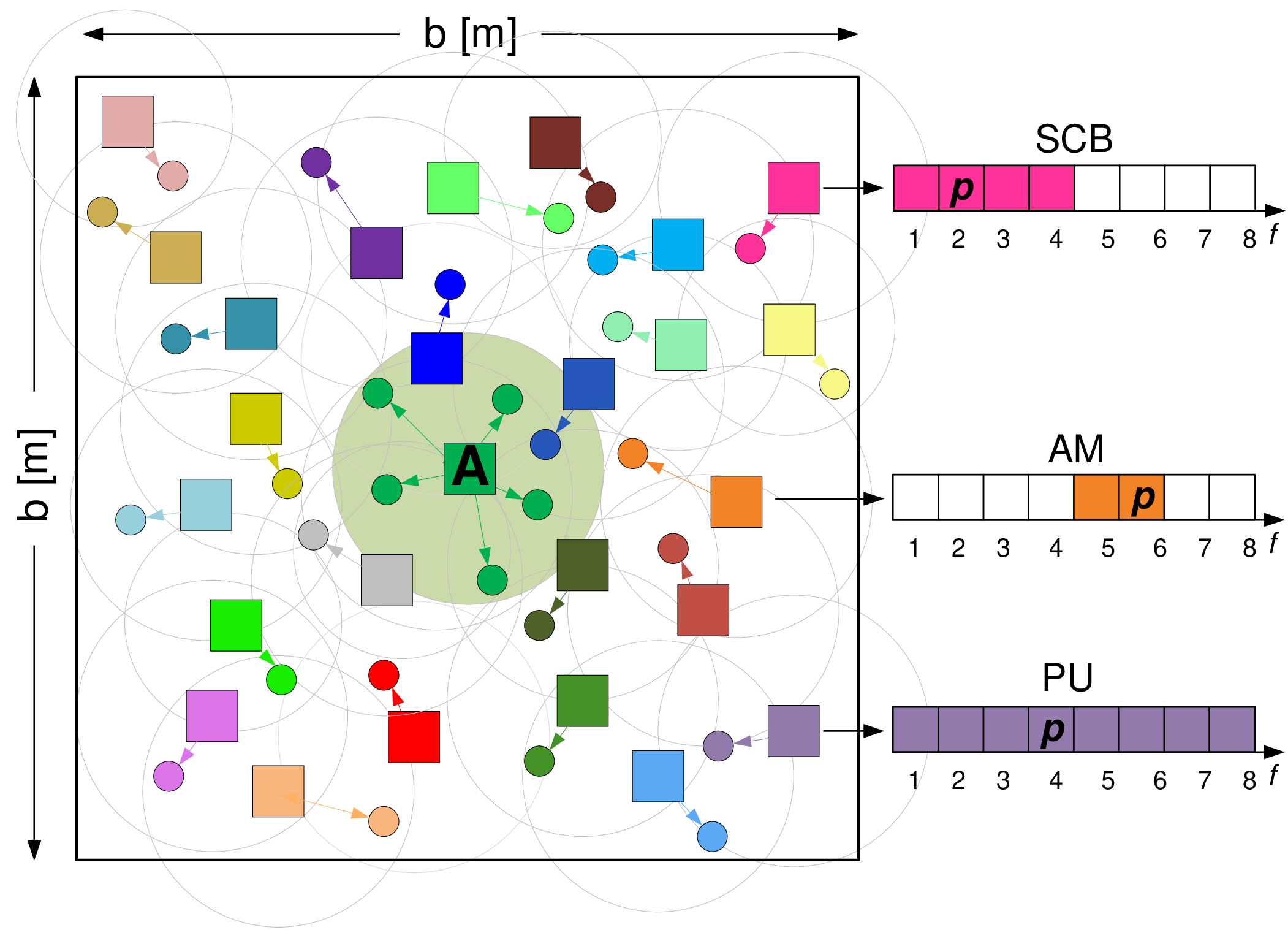}
		\caption{\newver{Central WLAN deployment with A placed in the middle and 24 WLANs spread uniformly at random.}}    
		\label{fig:map_central}
	\end{figure}

	% Scenario
	\newver{With the following experiment, we aim to identify what would be the optimal policy that a particular WLAN should adopt in order to increase its own throughput. In this case, we consider three rectangular maps (sparse, semi-dense and high-dense) with one WLAN (A) located at the center, and $M-1 = 24$ WLANs spread uniformly at random in the area (see Figure \ref{fig:map_central}). Besides, we now consider that WLAN A has $N_\text{A} \sim U[1,20]$ STAs.\footnote{\newverold{Note that the average results considering just one STA per WLAN are really similar to the ones presented in this work.}} Channel allocation (including the primary channel) is set uniformly at random to all the WLANs, except A. While the central WLAN is also set with a random primary channel, it is allocated the widest channel (i.e., $C_\text{A} = \{1,...,8\}$) in order to provide more flexibility and capture complex effects.}
	% Experiment procedure
	\newver{While the DCB policies of the rest $M-1$ WLANs are picked uniformly at random (i.e., they will implement OP, SCB, AM or PU with same probability 1/4), A's policy is set deterministically. Specifically, for each $N_b = 3$ map sizes (i.e., 75, 100 and 150 m$^2$), we generate $N_\text{D} = 400$ deployments following the aforementioned conditions for each of the DCB policies that A can implement. That is, we simulate $ N_\text{m} \times N_\text{D} \times N_\text{P} = 4800$ scenarios. The simulation time of each scenario is also $T_\text{obs} = 20$ seconds.}
	
	% Map figure

\begin{figure}[htbp]
		\small
		\centering
		\includegraphics[width=0.48\textwidth]{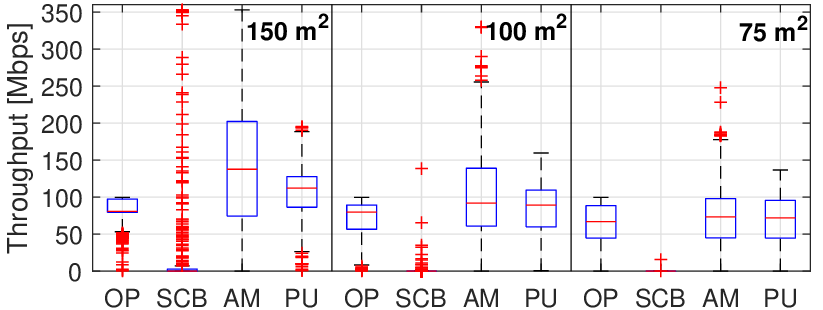}
		\caption{\newverold{DCB policy effect on central WLAN A.}}      
		\label{fig:throughput_boxplot_sim2}
	\end{figure}

	% Results
	% SCB is non-viable
	\newver{Figure \ref{fig:throughput_boxplot_sim2} shows the average throughput experienced by A in the considered maps. The first noticeable result is that, in dense scenarios, SCB is non-viable for WLANs with wide allocated channels because they are most likely prevented to initiate transmissions. In fact, A is not able to successfully transmit any data packets in 61\%, 84\% and 99\% of the scenarios simulated for SCB in the sparse, semi-dense and high-dense maps, respectively.}
	% Throughputs
	\newver{Regarding the rest of policies, on average, A's throughput is higher when implementing AM in all the maps. Especially, AM (and SCB in some cases) stands out in sparse deployment. Nevertheless, for dense deployments, there is a clear trend to pick just one channel when implementing AM or PU. That is why OP provides an average throughput relatively close to the ones achieved by these policies.}
	% Pie chart
	\newver{Nonetheless, as the high standard deviation of the throughput indicates, there are important differences regarding $\Gamma_\text{A}$ among the evaluated scenarios. Table \ref{table:results_central_pie_chart} compares the share of scenarios where AM or PU provide the highest individual throughput for A. We say that AM is better than PU if $\text{E}[\Gamma_\text{A}^\text{AM}] - \text{E}[\Gamma_\text{A}^\text{PU}] > \delta_\Gamma$, and vice versa. We use the margin $\delta_\Gamma = 1$ Mbps for capturing the cases where AM and PU perform similarly.}

	% Discuss pie chart
	\newver{We see that in most of the cases AM performs better than PU. However, in some scenarios, PU outperforms AM. This improvement is accentuated for the high-density map, where a significant 57\% (19\% + 38\%) of the scenarios achieve the same or highest throughput with PU. Also, there are a scenarios where the throughput experienced by PU with respect to AM is significantly higher (up to 61.5, 127.4 and 41.9 Mbps of improvement for the sparse, semi-dense and high-dense scenarios, respectively). This mainly occurs when the neighboring nodes occupy A's primary channel through complex interactions caused by information asymmetries, keeping its backoff counter frozen for long periods of time. These are clear cases where adaptive policies could importantly improve the performance.}
	
	% Conclusions
	\newver{Therefore, as a rule of thumb for dense networks, we can state that, while AM reaches higher throughput on average, stochastic DCB is less risky and performs relatively well. Nonetheless, even though PU is fairer than AM on average, it does not guarantee the absence of starving WLANs either. It follows that WLANs must be provided with some kind of adaptability to improve both the individual throughput and fairness with acceptable certainty.}
	
	\begin{table}[htbp]
		\small
		\centering
		\caption{Share of scenarios where AM or PU provide the highest individual throughput for WLAN A.}      
		\label{table:results_central_pie_chart}
		\begin{tabularx}{.49\textwidth}{CCCC}
			
			\toprule
			\textbf{Deployment} &
			\textbf{PU best} &
			\textbf{AM best} &
			\textbf{Draw}\\
			
			\midrule
			
			Sparse 150 m$^2$ & 90/400 (23\%) & 260/400 (65\%) & 50/400 (13\%) \\
			Semi 100 m$^2$ & 61/400 (15\%) & 246/400 (62\%) & 93/400 (23\%) \\
			High 75 m$^2$ & 77/400 (19\%) & 172/400 (43\%) & 151/400 (38\%) \\			
			
			\bottomrule
		\end{tabularx}
		
	\end{table}
	
	\section{Conclusions} \label{sec:conclusions}
	
	% What we have done in a phrase: DCB policies in spatially distributed scenarios
	\newver{In this work, we show the effects on spatially distributed WLANs of different DCB policies, including a new approach that stochastically selects the transmission channel width.}
	% SFCTMN analytical framework
	\newver{By means of modeling WLAN scenarios through CTMNs, we provide relevant insights such as the instantaneous vs. long-term throughput dilemma, i.e., always selecting the widest available channel found free does not always maximize the individual throughput. Besides, we show that often there is not an optimal global policy to be applied to each WLAN, but different policies are required, specially in non-fully overlapping scenarios where \textit{chain reaction} actions are complex to foresee.}
	
	% Komondor
	\newver{Simulations corroborate that, while AM is normally the optimal policy to maximize the individual long-term throughput, there are cases, particularly in \newver{high-density} scenarios, where stochastic DCB performs better both in terms of individual throughput and fairness among WLANs.}
	% Next steps
	\newver{We conclude that the performance of DCB can be significantly improved through adaptive policies capable of leveraging gathered knowledge from the medium and/or via information distribution. In this regard, our next works will focus on studying machine learning based policies to enhance WLANs spectrum utilization in high-density scenarios.}

	\section*{Acknowledgment}
	
	This  work  has  been  partially  supported  by  the  Spanish Ministry of Economy and Competitiveness under the Maria de Maeztu  Units  of  Excellence  Programme  (MDM-2015-0502),  and  by  a  Gift from the Cisco University Research Program (CG\#890107, Towards Deterministic Channel Access in High-Density WLANs) Fund, a corporate advised fund of Silicon Valley Community Foundation. The work done by S. Barrachina-Mu\~noz is supported by a FI grant from the Generalitat de Catalunya.
	
	\begin{appendices}

	\section{Dynamic channel bonding flowchart}
	\label{FirstAppendix}
	
\begin{figure}[h]
		\centering
		\includegraphics[width=0.48\textwidth]{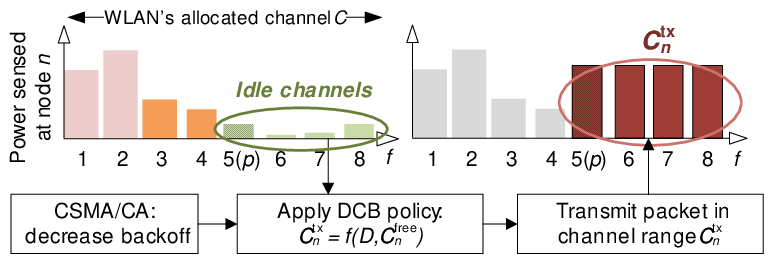}
		\caption{Flowchart of the transmission channel selection. In this example channel 5 is the primary channel and a DCB policy $\mathcal{D} = \text{AM}$ is applied.}    
		\label{fig:cb_policy_flowchart}
	\end{figure}

	In Figure \ref{fig:cb_policy_flowchart}, a simple flowchart of the transmission channel selection is shown.
	
	\section{Optimal combination of DCB policies}
	
	Table \ref{table:scenario_IV} shows the individual and aggregated throughputs, and the Jain's fairness index for all the combinations of DCB policies in \textit{Secnario IV}.
	
	\begin{table}[h]
		\small
		\centering
		\caption{Effect of different DCB policy combinations on throughput and fairness in the WLANs of \textit{Scenario IV}. The values obtained through \texttt{Komondor} are displayed in parentheses, while the other correspond to the SFCTMN framework.}
		\label{table:scenario_IV}
		\begin{tabularx}{.48\textwidth}{vvvCCCCCC}
			\toprule
			\multicolumn{3}{c}{\textbf{Policy}}	& \textbf{States}
			& \multicolumn{4}{c}{\textbf{Throughput [Mbps]}}	& \textbf{Fairness}		\\
			
			$\mathcal{D}_A$ & $\mathcal{D}_B$ & $\mathcal{D}_C$ & $\abs S$ & $\Gamma_\text{A}$  & $\Gamma_\text{B}$  & $\Gamma_\text{C}$  & $\Gamma$   & $\mathcal{J}$ \\
			
			\midrule
			
			% AM AM AM
			\multirow{2}{*}{AM}  & \multirow{2}{*}{AM}  & \multirow{2}{*}{AM}  & \multirow{2}{*}{5}  & 199.96     & 3.58   & 199.96 & \textbf{403.49}     &  0.67853   \\
			
			\multicolumn{1}{c}{} & & & & \scriptsize (199.35)     & \scriptsize (4.76)     & \scriptsize (199.37)	   &   \scriptsize (403.48)     & \scriptsize (0.68247)   \\ 
			
			% AM PU AM
			\multirow{2}{*}{AM}  & \multirow{2}{*}{PU}  & \multirow{2}{*}{AM}  & \multirow{2}{*}{10} & 149.41     & 62.45   & 149.41 & 361.27     & 0.89679   \\
			
			\multicolumn{1}{c}{} & & & & \scriptsize (128.72)   & \scriptsize (86.88)     & \scriptsize (128.72)	   &   \scriptsize (344.31)     & \scriptsize (0.97131)   \\ 
			
			% PU AM PU
			\multirow{2}{*}{PU}  & \multirow{2}{*}{AM}  & \multirow{2}{*}{PU}  & \multirow{2}{*}{14} & 109.84     & 108.44   & 109.84 & 328.12     & \textbf{0.99996}   \\
			
			\multicolumn{1}{c}{} & & & & \scriptsize (109.49)     & \scriptsize (109.13)     & \scriptsize (109.51)	   &   \scriptsize (328.13)          & \scriptsize (1.00000)   \\
			
			% AM AM PU
			\multirow{2}{*}{AM}  & \multirow{2}{*}{AM}  & \multirow{2}{*}{PU}  & \multirow{2}{*}{9} & 111.31     & 106.91   & 110.33 & 328.55         & \textbf{0.99970}   \\
			
			\multicolumn{1}{c}{} & & & & \scriptsize (109.64)     & \scriptsize (109.06)     &  (109.49)	   &   \scriptsize (328.19)     & \scriptsize (1.00000)   \\ 
			
			% AM PU PU
			\multirow{2}{*}{AM}  & \multirow{2}{*}{PU}  & \multirow{2}{*}{PU}  & \multirow{2}{*}{12} & 111.29     & 106.94   & 110.33     & 328.56     &  \textbf{0.99971}   \\
			
			\multicolumn{1}{c}{} & & & & \scriptsize (109.63)     & \scriptsize (109.07)     & \scriptsize (109.49)	   &   \scriptsize (328.18)     & \scriptsize (1.00000)   \\ 
			
			% PU PU PU
			\multirow{2}{*}{PU}  & \multirow{2}{*}{PU}  & \multirow{2}{*}{PU}  & \multirow{2}{*}{14} & 109.85     & 108.44   & 109.85 & 328.13     & \textbf{0.99996}  \\
			
			\multicolumn{1}{c}{} & & & & \scriptsize (109.52)     & \scriptsize (109.10)     & \scriptsize (109.51)	   &   \scriptsize (328.13)     & \scriptsize (1.00000)   \\ 
			
			\bottomrule
		\end{tabularx}
	\end{table}
	
	\section{Evaluation setup}
	% Table of parameters
	\newverold{The values of the parameters considered in the simulations are shown in Table \ref{table:appendix_table}. % Path loss
		Regarding the path loss, we use the dual-slope log-distance model for 5.25 GHz indoor environments in room-corridor condition \cite{xu2007indoor}. Specifically, the path loss in dB experienced at a distance $d$ is defined by
		\begin{equation}	\label{eq:path_loss}
		\text{PL}(d) = 
		\left\{
		\begin{array}{ll}
		53.2 \Plus 25.8 \log_{10}(d)  & \mbox{if } d \leq d_1 \text{ m} \\
		56.4 \Plus 29.1 \log_{10}(d)  & \mbox{otherwise}
		\end{array}
		\right.\text{,}
		\end{equation} 
		where $d_1 = 9$ m is the break point distance.}
	
	\begin{table}[h]
		\small
		\caption{Parameters considered in the presented scenarios.}
		\label{table:appendix_table}
		\centering
		\begin{tabularx}{.47\textwidth}{vbv}
			\toprule
			\textbf{Parameter}     & \textbf{Description}              & \textbf{Value} \\ 
			% PHY and scenario config
			\midrule
			$f_\text{c}$ & Central frequency           & 5 GHz  \\
			$\abs c$ & Basic channel bandwidth          & 20 MHz \\
			$L_\text{D}$       & Frame size           & 12000 bits     \\ 
			$N_\text{a}$       & No. of frames in an A-MPDU & 64             \\  
			$\text{CW}_\text{min}$ & Min. contention window            & 16             \\ 
			$m$                    & No. of backoff stages          & 5              \\   
			MCS						& 11ax MCS index							& 0 - 11		\\
			$\eta$                 & MCS's packet error rate        & 0.1           \\ 
			CCA                    & CCA threshold                               & -82 dBm        \\ 
			$P_\text{tx}$          & Transmission power                & 15 dBm         \\ 
			$G_\text{tx}$         & Transmitting gain                 & 0 dB           \\ 
			$G_\text{rx}$         & Reception gain                    & 0 dB           \\ 
			$\text{PL}(d)$		& Path loss 			& see (\ref{eq:path_loss})		\\
			$\text{P}_\nu$ & Adjacent power leakage factor        & -20 dB  \\
			CE                     & Capture effect threshold          & 20 dB          \\ 
			$N$                      & Background noise level            & -95 dBm        \\
			% MAC
			\midrule
			$T_\text{e}$       & Empty backoff slot duration                     & 9 \textmu s          \\
			$T_\text{SIFS}$                   & SIFS duration                     & 16 \textmu s      \\ 
			$T_\text{DIFS}$                   & DIFS duration                     & 34 \textmu s      \\ 
			$T_\text{PIFS}$                   & PIFS duration                     & 25 \textmu s      \\
			$T_\text{PHY-leg}$      & Legacy preamble     & 20 \textmu s           \\ 
			$T_\text{PHY-HE-SU}$      & HE single-user preamble       & 164 \textmu s \\
			$\sigma_\text{leg}$      & Legacy OFDM symbol duration     & 4 \textmu s           \\
			$\sigma$      & 11ax OFDM symbol duration     & 16 \textmu s           \\
			$L_\text{BACK}$       & Length of a block ACK             & 432 bits       \\ 
			$L_\text{RTS}$        & Length of an RTS packet           & 160 bits       \\ 
			$L_\text{CTS}$        & Length of a CTS packet            & 112 bits       \\ 
			$L_\text{SF}$      & Length of service field       & 16 bits           \\ 
			$L_\text{MD}$      & Length of MPDU delimiter       & 32 bits           \\ 
			$L_\text{MH}$      & Length of MAC header     & 320 bits           \\ 
			$L_\text{TB}$      & Length of tail bits     & 18 bits           \\ 
			\bottomrule
		\end{tabularx}
	\end{table}
	% MCS and data rate
	\newverold{The MCS index used for each possible channel bandwidth (i.e., 20, 40, 80 or 160 MHz) was the highest allowed according to \textit{i}) the power power budget established between the WLANs and their corresponding STA/s, and \textit{ii}) the minimum sensitivity required by the MCSs. As stated by the 11ax amendment, the number of transmitted bits per OFDM symbol used in data transmissions is given by the channel bandwidth and the MCS parameters, i.e., $r = Y_\text{sc} Y_\text{m} Y_\text{c} V_\text{s}$, where $Y_\text{sc}$ is the number of data sub-carriers, $Y_\text{m}$ is the number of bits in a modulation symbol, $Y_\text{c}$ is the coding rate, and $ V_\text{s} = 1$ is the number of single user spatial streams (note that we only consider one stream per transmission).}
	
	\newverold{The number of data sub-carriers depends on the transmission channel bandwidth. Specifically, $Y_\text{sc}$ can be 234, 468, 980 or 1960 for 20, 40, 80, and 160 MHz, respectively. For instance, the data rate provided by MCS 11 in a 20 MHz transmission is $s = (234 \times 10 \times 5/6 \times 1)\sigma^{-1} = 121.9$ Mbps.
		% Control frames
		However, control frames are transmitted in legacy mode using the basic rate $r_\text{leg} = 24$ bits per OFDM symbol of MCS 0, corresponding to $s_\text{leg} = 6$ Mbps since the legacy OFDM symbol duration $\sigma_\text{leg}$ must be considered. With such parameters we can define the duration of the different packets transmissions, and the duration of a successful and collision transmission accordingly:
		\begin{flalign*}
		T_\text{RTS} &= T_\text{PHY-leg} \Plus \ceil*{\frac{L_\text{SF} \Plus L_\text{RTS} \Plus L_\text{TB}}{r_\text{leg}}} \sigma_\text{leg}  \text{,} \\
		T_\text{CTS} &= T_\text{PHY-leg} \Plus \ceil*{\frac{L_\text{SF} \Plus L_\text{CTS} \Plus L_\text{TB}}{r_\text{leg}}} \sigma_\text{leg} \text{,} \\
		T_\text{DATA} &= T_\text{PHY-HE-SU} \Plus \\
		& \Plus \ceil*{\frac{L_\text{SF} \Plus N_\text{a} (L_\text{MD} \Plus L_\text{MH} \Plus L_\text{D})  \Plus L_\text{TB}}{r}} \sigma \text{,} \\
		T_\text{BACK} &= T_\text{PHY-leg} \Plus \ceil*{\frac{L_\text{SF} \Plus L_\text{BACK} \Plus L_\text{TB}}{r_\text{leg}}} \sigma_\text{leg} \text{.}
		\end{flalign*}}
	
\end{appendices}

	\bibliographystyle{unsrt}
	\bibliography{bib}
		
	\vfill
	
	% that's all folks
\end{document}